\documentclass[aps,pra,10pt,nobalancelastpage,twocolumn,nofootinbib,superscriptaddress,floatfix]{revtex4-2}
\usepackage{xcolor,amsthm,amsmath,amsxtra,amsfonts,dsfont,graphicx,bm,comment,braket}
\usepackage[normalem]{ulem}
\usepackage[colorlinks,citecolor=blue,linkcolor=magenta,urlcolor=blue,breaklinks=true]{hyperref}

\newcommand{\tr}{\mathrm{tr}}
\newcommand*\dd{\mathop{}\!\mathrm{d}}
\newcommand{\Pmodel}{\mathbb{P}_{\text{model}}}
\newcommand{\Pdata}{\mathbb{P}_{\text{data}}}

\newtheorem*{problem*}{Problem}

\begin{document}

\title{Tensor Networks and Efficient Descriptions of Classical Data}

\author{Sirui Lu}
\email{sirui.lu@mpq.mpg.de}
\author{M\'arton Kan\'asz-Nagy}
\author{Ivan Kukuljan}
\author{J. Ignacio Cirac}
\affiliation{Max-Planck-Institut f\"ur Quantenoptik, Hans-Kopfermann-Str.\ 1, 85748 Garching, Germany}
\affiliation{Munich Center for Quantum Science and Technology (MCQST), Schellingstr.\ 4, D-80799 M\"unchen, Germany}

\begin{abstract}
    We investigate the potential of tensor network-based machine learning methods to scale to large image and text datasets. We study how the mutual information between a subregion and its complement scales with the subsystem size $L$, similarly to approaches used in quantum many-body physics.
    Using various MI estimation methods, including a novel autoregressive network-based estimator, we find that simple image datasets exhibit area law scaling, suggesting efficient representation by two-dimensional tensor networks. More complex image datasets exceed the area law, indicating the need for generalized tensor network states or hybrid models. For Wikipedia text data, we observe power-law scaling $I(L) \propto L^\nu$ with $\nu \approx 0.82$, approaching a volume law. This implies that one-dimensional tensor networks with area law entanglement may not efficiently capture the structure of text. We introduce two models to reproduce this scaling: a quantum-inspired random pair toy model, and a linguistically-motivated Markovian dependency tree model. In the latter model, matching the observed MI scaling allows us to infer the word-word correlation length distribution in text.
\end{abstract}

\date{\today}

\maketitle

\section{Introduction}
\label{introduction}

The past decade has witnessed remarkable advancements in machine learning, primarily driven by deep learning techniques~\cite{Lecun2015Deep,Jordan2015Machine}. Despite impressive practical successes in tasks such as text and image classification~\cite{Krizhevsky2012Imagenet,Kim2014Convolutional}, generation~\cite{PixelRCNN,GAN,VanDenOord2016Conditional}, and representation learning~\cite{Bengio2013Representation}, the theoretical foundations of deep learning remain an active area of research. Perspectives from information theory~\cite{Tishby2015Deep,Cohen2015Expressive,Levine2018Deep,Levine2018Quantum,Zhang2017Entanglement,Huang2017Provably}, statistical mechanics~\cite{Bahri2020Statistical}, and renormalization~\cite{Beny2013Deep,Mehta2014Exact,Gallego2022Language,Koch-Janusz2018Mutual} are still being developed. A central question is how neural networks capture the relevant corners of the data space occupied by natural images and text~\cite{Lin2017Why}.

In contrast, quantum many-body physics offers a mature theoretical framework for understanding quantum data and identifying suitable representations. Tensor networks, supported by rigorous theory based on entanglement~\cite{Eisert2010Colloquium} and mutual information (MI)~\cite{Wolf2008Area}, provide an efficient representation for quantum data. Analogous to how neural networks capture the manifold of natural images and text, tensor networks efficiently represent low-energy states of quantum many-body systems~\cite{Cirac2021Matrix}. These states, while complex, occupy only a small corner of the exponentially large Hilbert space. Identifying this corner is essential for developing efficient numerical methods~\cite{Schollwock2011Densitymatrix}.

\begin{figure}
    \includegraphics[width=0.99\linewidth]{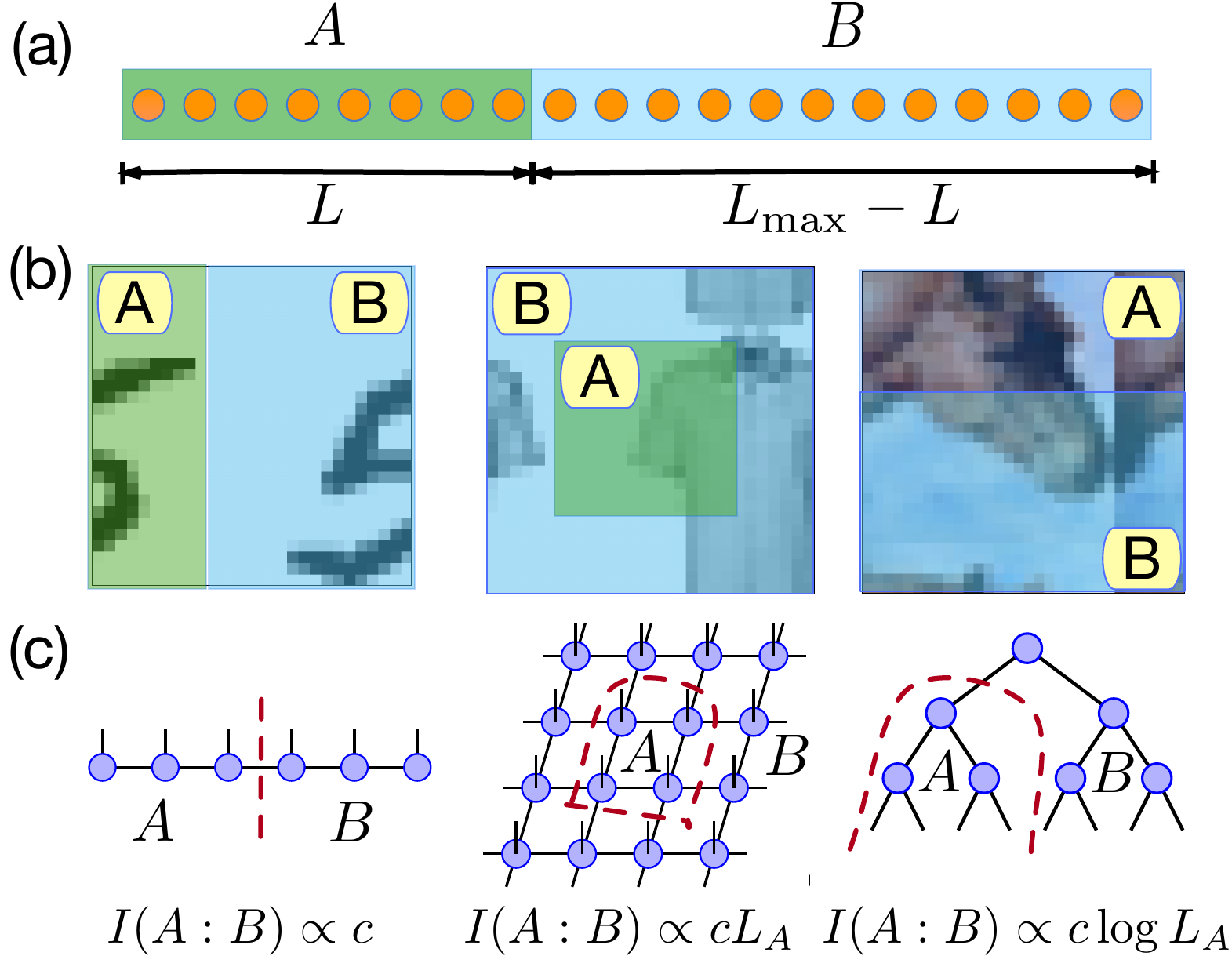}
    \caption{(a) Schematic illustration of one-dimensional partitioning into regions $A$ and $B$ of lengths $L$ and $L_{\max}-L$, where $L_{\max}$ is the total system size. (b) Three distinct partitioning schemes for two-dimensional image data: left:right ($\text{L}:\text{R}$), center:surroundings ($\text{C}:\text{S}$), and top:bottom ($\text{T}:\text{B}$). Example images are from MNIST (handwritten digit 5), Fashion-MNIST (T-shirt), and CIFAR-10 (horse head) datasets, with random displacements applied to ensure translational invariance. (c) Tensor network representations and their characteristic mutual information scaling: matrix product states (MPS) exhibit constant scaling $I(A:B) \propto c$, projected entangled pair states (PEPS) show linear scaling with boundary length $I(A:B) \propto cL_A$, and tree tensor network states display logarithmic scaling $I(A:B) \propto c\log L_A$. When area laws hold, the mutual information $I(A:B)$ is bounded by the number of sites near the $A$-$B$ boundary.}
    \label{fig:A-schematic-illustration}
\end{figure}

Characterizing low-energy states through entanglement or MI scaling has been instrumental in identifying suitable variational quantum states. For example, in one-dimensional systems, it is rigorously proven that the scaling of entanglement between a subsystem and the rest of the system must align with that of the tensor network states used to represent it. Otherwise, tensor network approximations become computationally intractable~\cite{Schuch2008Entropy}. This can be explained in terms of the area law for entanglement: states satisfying an area law can be efficiently represented using matrix product states (MPS), a one-dimensional family of tensor networks (TNs), as depicted in Fig.~\ref{fig:A-schematic-illustration}(c). In contrast, for quantum critical systems, where the entanglement entropy scales logarithmically with the subsystem size, an MPS description requires a bond dimension that grows polynomially with the system size~\cite{Verstraete2006Matrix}. In such cases, multilayer TNs such as the multi-scale entanglement renormalization ansatz (MERA)~\cite{Vidal2008Class} or tree tensor network (TTN) structures~\cite{Shi2006Classical}, shown in Fig.~\ref{fig:A-schematic-illustration}(c), are more suitable since they also exhibit critical scaling of entanglement entropy. This understanding also allows one to rule out certain architectures: for instance, TNs are not the appropriate choice for out-of-equilibrium systems that develop volume law correlations over long times~\cite{Schuch2008Entropy}.

Given their success in quantum physics and connections to probabilistic graphical models~\cite{Robeva2019Duality,Glasser2020Probabilistic} of machine learning, tensor networks have recently gained attention in classical machine learning. They have found applications in unsupervised~\cite{Han2018Unsupervised,Stoudenmire2018Learning,Glasser2019Expressive,Cheng2019Tree,Efthymiou2019TensorNetwork,Huggins2019Quantum,Miller2021Tensor,Cheng2021Supervised} and supervised learning tasks~\cite{Stoudenmire2018Learning,Carleo2019Machine,Glasser2019Expressive,Glasser2020Probabilistic,Cheng2021Supervised}, as well as in compressing neural networks~\cite{Novikov2015Tensorizing,Gao2020Compressing}. Empirical studies suggest that using tensor networks can significantly reduce computational resources and speed up processes in text generation~\cite{Novikov2015Tensorizing} and image processing~\cite{Gao2020Compressing}. However, a critical question remains: Are tensor networks capable of providing an adequate and efficient representation of real-world data distributions?

In this work, we investigate the scaling of mutual information in text and images, a measure that parallels quantum entanglement in classical datasets~\cite{Wilms2011Mutual}. Our analysis reveals that natural text exhibits power-law scaling of MI between subsets and their complements, with an exponent $\nu \approx 0.82$, approaching a volume law. This suggests that traditional one-dimensional tensor network approaches, such as MPS or tree tensor networks (TTN), may not scale efficiently to long texts. Surprisingly, despite the known presence of power-law decaying correlations in natural text~\cite{Lin2017Critical,Shen2019Mutual}, the MI scaling in text approaches a volume law rather than the logarithmic scaling seen in quantum critical systems--a scaling that the hidden Markov tree model introduced by Lin and Tegmark~\cite{Lin2017Critical} would predict, according to our analysis based on the connection between TNs and graphical models. To reconcile this discrepancy, we introduce two toy models: a quantum-inspired random pair model~\cite{Wolf2008Area}, which shows that algebraic correlations in classical probability distributions can coexist with power-law mutual information scaling, and a Markov generative model based on dependency parsing trees~\cite{Chomsky1957Syntactic,Kubler2009Dependency} from natural language processing that incorporates linguistic dependencies. This refined model also reproduces the observed MI scaling when dependency lengths between words follow a power-law distribution.

For image data, we examine widely used datasets such as MNIST~\cite{mnist}, Fashion-MNIST~\cite{Xiao2017Fashionmnist}, and CIFAR-10~\cite{cifar} (see Fig.~\ref{fig:A-schematic-illustration}(b)). Area law scaling in these datasets would imply that two-dimensional tensor networks like projected entangled pair states (PEPS)~\cite{Verstraete2004Renormalization}, with moderately growing bond dimensions, could efficiently represent them. PEPS have proven effective in representing two-dimensional gapped local systems, which also follow area law behavior~\cite{Murg2007Variational,Corboz2011Stripes,Bauer2012Threesublattice,Corboz2016Variational,Rader2018Finite,Ponsioen2019Period,Vanhecke2022Scaling,Vlaar2021Simulation}. Previous studies~\cite{Cheng2017Information,Martyn2020Entanglement,Convy2022Mutual} have reported somewhat conflicting results on MI scaling in MNIST. While Ref.~\cite{Martyn2020Entanglement} suggested scaling stronger than the area law, Ref.~\cite{Cheng2017Information} found no definitive evidence for area law scaling in MNIST. A recent study~\cite{Convy2022Mutual} reported area law scaling in the TinyImages dataset~\cite{Torralba200880}\footnote{After completing the work reported in the first arXiv version, we became aware of a related study~\cite{Convy2022Mutual}, focusing on MI scaling in image data. This work finds area-law scaling in the Tiny Images dataset~\cite{Torralba200880}, but not in MNIST. We attribute the discrepancy with our work to our approach of making datasets translationally invariant to avoid featureless outer regions in MNIST. Compared to the initial version, the current version also includes the dependency tree model.}, but not in MNIST. Our analysis, based on various MI estimation methods, shows close-to-area-law scaling for the simpler MNIST dataset, which consists of handwritten images. This supports the potential of PEPS-like tensor networks for representing MNIST~\cite{Cheng2021Supervised}. However, our results for more complex datasets like Fashion-MNIST and CIFAR-10 indicate scaling beyond the area law.

Estimating mutual information from classical datasets is an active research area~\cite{Grassberger2003Entropy,Kraskov2004Estimating,Belghazi2018MINE,Song2020Understanding,Carrara2019Estimation}, with many current methods struggling with scalability and stability due to high-dimensional probability distributions. In this work, we develop an MI estimator for images based on an autoregressive network model~\cite{PixelRCNN} and benchmark it against a standard $k$-nearest neighbor (kNN) density estimation method~\cite{Kraskov2004Estimating}. We also utilize the mutual information neural estimator (MINE)~\cite{Belghazi2018MINE}, enhancing it by using convolutional neural networks (CNNs) as variational functions. By increasing network complexity--from simple fully connected networks to CNNs--we systematically improve MI estimation accuracy. Despite previous challenges in training MINE estimators~\cite{McAllester2018Formal}, we achieve stable and consistent results across image and text data.

The remainder of this paper is organized as follows: In Sec.~\ref{sec:TN_and_ML}, we discuss how entanglement entropy and mutual information scaling empower or constrain the applicability of tensor network models for quantum states and classical data. In Sec.~\ref{sec:MI_est}, we present our numerical MI estimation methods, including our autoregressive network-based MI estimator, MINE with CNNs, and the kNN estimator as a reference. In Secs.~\ref{sec:MI_image} and~\ref{sec:MI_text}, we analyze the mutual information scaling in images and text numerically and introduce toy models to interpret the power-law scaling of both MI and correlations in text. Finally, we summarize our findings and discuss their implications for future research in Sec.~\ref{sec:conclusion}.

\section{Efficient Description of Quantum Systems and Classical Data}
\label{sec:TN_and_ML}

\subsection{Entanglement, Tensor Networks, and Quantum Many-Body Systems}

Understanding quantum many-body systems on a lattice requires grappling with the exponential growth of the Hilbert space with the number of lattice sites, a consequence of the tensor product structure of local Hilbert spaces. This challenge mirrors the difficulties in machine learning with large datasets: the space is too vast to fully process or store. In quantum physics, the resolution relies on the concept of entanglement entropy. Low-energy states of local Hamiltonians, particularly ground states, occupy a small subset of the full Hilbert space characterized by low entanglement entropy. This allows for their efficient numerical representation through tensor network states that reflect this entanglement scaling~\cite{Cirac2021Matrix,Schollwock2011Densitymatrix}.

To clarify what is meant by ``low entanglement'', consider a quantum state described by the density matrix $\rho_{AB}$. For a subset $A$ of a system and its complement $B$ (see Fig.~\ref{fig:A-schematic-illustration}), the reduced density matrix $\rho_A = \tr_B(\rho_{AB})$ captures the state of $A$, and similarly for $B$. When the density matrix is pure, i.e., $\rho_{AB}=\ket{\Psi}\bra{\Psi}$ for some $\Psi$, the entanglement entropy (EE) is defined as
\begin{equation}
    S(\Psi) = S(\rho_A) = -\tr(\rho_A \log \rho_A) = S(\rho_B),
\end{equation}
measuring the entanglement between $A$ and $B$. The scaling of EE with the size of $A$ is of significant interest in quantum many-body physics. Systems with finite correlation lengths obey an area law for EE, where $S(\rho_A) \propto |\partial A|$, growing with the boundary of $A$ rather than its volume~\cite{Wolf2008Area}. This has been rigorously established for one-dimensional gapped systems with local Hamiltonians~\cite{Hastings2007Area} and for two-dimensional systems under certain conditions~\cite{Anshu2020Entanglement,Verstraete2013EntRate,Marien2016Entanglement}. In contrast, gapless systems like free theories, critical systems, and conformal field theories (CFTs) exhibit diverging correlation lengths and logarithmic EE scaling in one dimension, $S(\rho_A) \propto \log|A|$, and area law scaling $S(\rho_A) \propto |A|$ in two dimensions~\cite{Srednicki1993Entropy,Wilczek1994Geometric,Vidal-Rico-Latorre-Kitaev,Calabrese2004Entanglement,CalabreseCardy2009,CasiniHuerta2009}.

The logarithmic or area law scaling of EE in ground states suggests that these states only occupy a restricted region of the Hilbert space, making them amenable to efficient characterization through variational states with matching entanglement scaling~\cite{Wolf2008Area,Schuch2008Entropy,Cirac2021Matrix}. Tensor networks are particularly effective at representing ground states of many-body local Hamiltonians~\cite{Cirac2021Matrix}. In one dimension, matrix product states (MPS) comply with the area law and serve as effective variational states for gapped Hamiltonians~\cite{Wolf2008Area}. The EE of multi-scale entanglement renormalization ansatz (MERA) and tree tensor networks (TTN) scales logarithmically, aligning with the logarithmic EE scaling of quantum critical systems~\cite{Vidal2008Class}. In higher dimensions, projected entangled pair states (PEPS) and MERA follow the area law, while TTN exhibit logarithmic scaling~\cite{Cirac2021Matrix}.

Empirical evidence and theoretical proofs demonstrate that tensor network methods can approximate states with desired accuracy when their entanglement scaling matches that of the tensor network. For instance, one-dimensional systems with area law entanglement can be efficiently represented by MPS with constant bond dimensions~\cite{Verstraete2006Matrix}. States with logarithmic entanglement corrections can also be described with MPS with polynomially growing bond dimensions, though MERA and TTN may be more suitable for certain models. Moreover, algorithms based on MPS for ground states of one-dimensional gapped Hamiltonians are known to be efficient~\cite{Landau2015Polynomial}. In higher dimensions, PEPS and MERA are proven to follow area laws, making them suitable for representing ground states of gapped local systems~\cite{Wolf2008Area}.

However, certain states violate the area law for entanglement. Notably, non-equilibrium states arising from long-time evolution under local Hamiltonians often exhibit entanglement entropy that grows linearly with time, $S(t) \propto t$~\cite{Calabrese2004Entanglement,Schuch2008Entropy,Schuch2008Entropya}. Such states eventually require exponentially large bond dimensions for accurate MPS representations, rendering them inaccurately describable by tensor networks.

\subsection{Mutual Information, Generative Models, and Classical Data}

Mutual information (MI) is a key information-theoretic measure for quantifying interdependence between variables, applicable to both classical and quantum systems~\cite{Cover2006Elements,Wolf2008Area}. For mixed states like thermal states, which are quantum generalizations of classical Boltzmann distributions, the entanglement entropy (EE) accounts not only for quantum entanglement but also the degree of mixedness and thermal entropy, leading to mixed signals that cannot be diagnosed without looking at MI. The mutual information between subsystems $A$ and $B$ is defined as:
\begin{equation}
    I(A:B) = S(\rho_A) + S(\rho_B) - S(\rho_{AB}),
    \label{eq:defMutInfoEntropy}
\end{equation}
where $S(\rho)$ denotes the entropy of the state $\rho$.
MI is advantageous as it provides an upper bound for all correlation functions, and it has been shown that thermal states with finite interaction ranges follow an area law for MI scaling~\cite{Wolf2008Area}. Indeed, such thermal states can be efficiently represented using tensor networks~\cite{Hastings2006Solving,Molnar2015Approximating}.

For classical data, MI is defined using Shannon entropy:
\begin{equation}
    S(A) = -\int_{\mathcal{A}} \mathbb{P}_A(a) \log \mathbb{P}_A(a) \,\dd a,
\end{equation}
where $\mathbb{P}_{AB}$ is the joint probability distribution of the data, and $\mathbb{P}_{A} = \int_{\mathcal{B}} \mathbb{P}_{AB}(a,b)\,\dd b$ is the marginal distribution of subsystem $A$. MI then quantifies the information shared between $A$ and $B$.

In classical data contexts, generative models play a role analogous to tensor networks for quantum many-body states. These models aim to replicate the probability distribution of datasets, such as images and text. Given a dataset $\mathcal{D}$ with $M$ samples $\{\boldsymbol{x}_1,\ldots,\boldsymbol{x}_M\}$ sampled from the true distribution $\Pdata(\boldsymbol{x})$, the task of generative modeling is to construct a model distribution $\Pmodel(\boldsymbol{x})$ capable of generating new samples resembling the original data.

Generative models fall into two categories: explicit density models and implicit density models. Explicit density models directly learn $\Pmodel(\boldsymbol{x})$ and allow computation of probabilities for any input $\boldsymbol{x}$. Examples include probabilistic graphical models~\cite{Bishop2006Pattern}, autoregressive neural networks~\cite{PixelRCNN,VanDenOord2016Conditional,Uria2016Neural,Salimans2017PixelCNN}, and sequence models in natural language processing~\cite{Sutskever2011Generating}. Some tensor networks, known in machine learning as Tensor Trains~\cite{Oseledets2011TensorTrain} and Tensor Trees, also belong to this category and benefit from efficient contraction properties. Implicit density models, such as Boltzmann machines~\cite{Salakhutdinov2010Efficient} and generative adversarial networks (GANs)~\cite{GAN}, generate data samples without explicitly computing probabilities. Although PEPS are theoretically explicit density models, the computational complexity of contracting them exactly (which is~\#P-Complete~\cite{Schuch2007Computational}) necessitates approximate contraction methods, categorizing them as implicit models in practice~\cite{Lubasch2014Unifying}. Both model types have demonstrated success across various tasks. Explicit models are particularly valuable when the log-likelihood needs to be computed, whereas implicit models excel in high-quality sample generation.

Drawing from experiences in quantum many-body physics, for a model (whether a neural network or tensor network) to effectively learn a dataset, it must be capable of generating probability distributions with MI scaling that is at least as rapid as that of the actual data. Thus, we investigate the MI scaling behavior in classical data to assess the scalability of tensor networks as generative models for large-scale machine learning tasks.

As illustrated in Fig.~\ref{fig:A-schematic-illustration}, we partition the data into two subsystems, $A$ and $B$, to study their mutual information $I(A:B)$. For one-dimensional data like text, we use a left-right partition. For two-dimensional data, such as images, we examine two types of separations. The first is a horizontal cut, dividing the system into top-bottom ($\text{T}:\text{B}$) regions, with $L$ representing the length of the top region. The second is a center-surroundings partition ($\text{C}:\text{S}$), where $L$ is the side length of the central square.

If the area law holds, the mutual information between two regions should be proportional to the interface area between them. For one-dimensional text data, this implies a constant MI, $I(\text{L}:\text{R}) = \text{constant}$, provided $A$ and $B$ are sufficiently large. For two-dimensional image data, $I(\text{T}:\text{B})$ should remain constant for top-bottom partitions, while for center-surroundings partitions, the interface area increases linearly with $L$, suggesting $I(\text{C}:\text{S}) \sim L$. In finite systems, MI is expected to grow non-linearly before stabilizing at the area law plateau, as MI is zero at the boundaries. In the case of volume laws, the MI between the two regions grows with the volume of the smaller region. This occurs, e.g., when each part of the system is correlated with every other part. Initially, for one region very small and the other large, this yields $I(\text{L}:\text{R}) \sim L$, and for image data, $I(\text{T}:\text{B}) \sim L$ and $I(\text{C}:\text{S}) \sim L^2$. Intermediate MI scaling behaviors, such as logarithmic $I(A:B) \sim \log(L)$ or power-law $I(A:B) \sim L^\alpha$ for $0 < \alpha < 1$, are also possible.

\section{Estimators of Mutual Information}
\label{sec:MI_est}

Estimating mutual information (MI) from empirical data is crucial for uncovering complex relationships within datasets. This problem can be formally stated as:
\begin{problem*}[Mutual information estimation from samples]
    Given $N$ independent and identically distributed (i.i.d.) samples $(a_i,b_i)$, $i=1,\ldots,N$, from the joint probability density $\mathbb{P}_{AB}$, estimate the mutual information $I(A:B)$ defined as
    \begin{equation}
        \label{eq:defMI}
        I(A: B)=\int_{\mathcal{A} \times \mathcal{B}} \mathbb{P}_{AB}(a,b) \log \frac{\mathbb{P}_{AB}(a,b)}{\mathbb{P}_{A}(a)\mathbb{P}_{B}(b)}\,\dd a\,\dd b,
    \end{equation}
    where $\mathbb{P}_{A}(a)=\int_{\mathcal{B}} \mathbb{P}_{AB}(a,b)\,\dd b$ and $\mathbb{P}_{B}(b)=\int_{\mathcal{A}} \mathbb{P}_{AB}(a,b)\,\dd a$ are the marginal probability densities of $A$ and $B$, respectively.
\end{problem*}
This task is central to various fields, including deep learning and information theory~\cite{Tishby2015Deep,DevonHjelm2019Learning}. However, accurately determining MI from high-dimensional data remains challenging due to the curse of dimensionality.

Estimation methods are broadly classified into parametric and nonparametric approaches~\cite{Paninski2003Estimation}. Nonparametric methods do not assume a specific data distribution model but often struggle with high-dimensional data. Parametric methods, conversely, use model-based approaches with adjustable parameters to approximate the underlying distributions.

In this work, we employ both parametric and nonparametric methods to estimate MI, leveraging their respective strengths and cross-validating their results. We introduce an MI estimator using density estimates provided by advanced autoregressive neural networks~\cite{Gregor2014Deep,Uria2016Neural,Reed2017Parallel}. We implement the mutual information neural estimator (MINE)~\cite{Belghazi2018MINE} enhanced by convolutional neural networks. To complement these parametric methods, we use the standard $k$-nearest neighbor (kNN) estimator as a nonparametric benchmark. This section provides an overview of these estimators, with detailed implementations discussed in Appendices~\ref{App:kNN} and~\ref{App:MINE}.

\subsection{Estimation from Trained Autoregressive Networks}
\label{sec:AutoregressiveNetworks}

We propose using autoregressive neural networks, a tractable explicit density model, to estimate entropy and MI. Autoregressive models~\cite{Uria2016Neural,PixelRCNN,VanDenOord2016Conditional,Salimans2017PixelCNN} decompose the joint probability distribution into a product of conditional probabilities:
\begin{equation}
    \mathbb{P}(\boldsymbol{x})=\prod_i \mathbb{P}(x_i|\boldsymbol{x}_{<i}),
    \label{eq:pARN}
\end{equation}
where $\boldsymbol{x}_{< i}=[x_1, x_2, \ldots, x_{i-1}]$ represents the vector of variables preceding $x_i$. These conditional probabilities are defined as parameterized functions with a fixed number of parameters. We consider the conditional distributions $\mathbb{P}(x_i|\boldsymbol{x}_{<i})$ as Bernoulli random variables, defined by a function (to be learned) that maps $\boldsymbol{x}_{<i}$ to the mean of the Bernoulli distribution. Popular architectures for autoregressive models include WaveNet~\cite{Oord2016WaveNet}, PixelRNN~\cite{PixelRCNN}, PixelCNN~\cite{VanDenOord2016Conditional}, and PixelCNN++~\cite{Salimans2017PixelCNN}, which have demonstrated excellent performance in various tasks.

\begin{figure}[ht]
    \centering
    \includegraphics[width=0.9\linewidth]{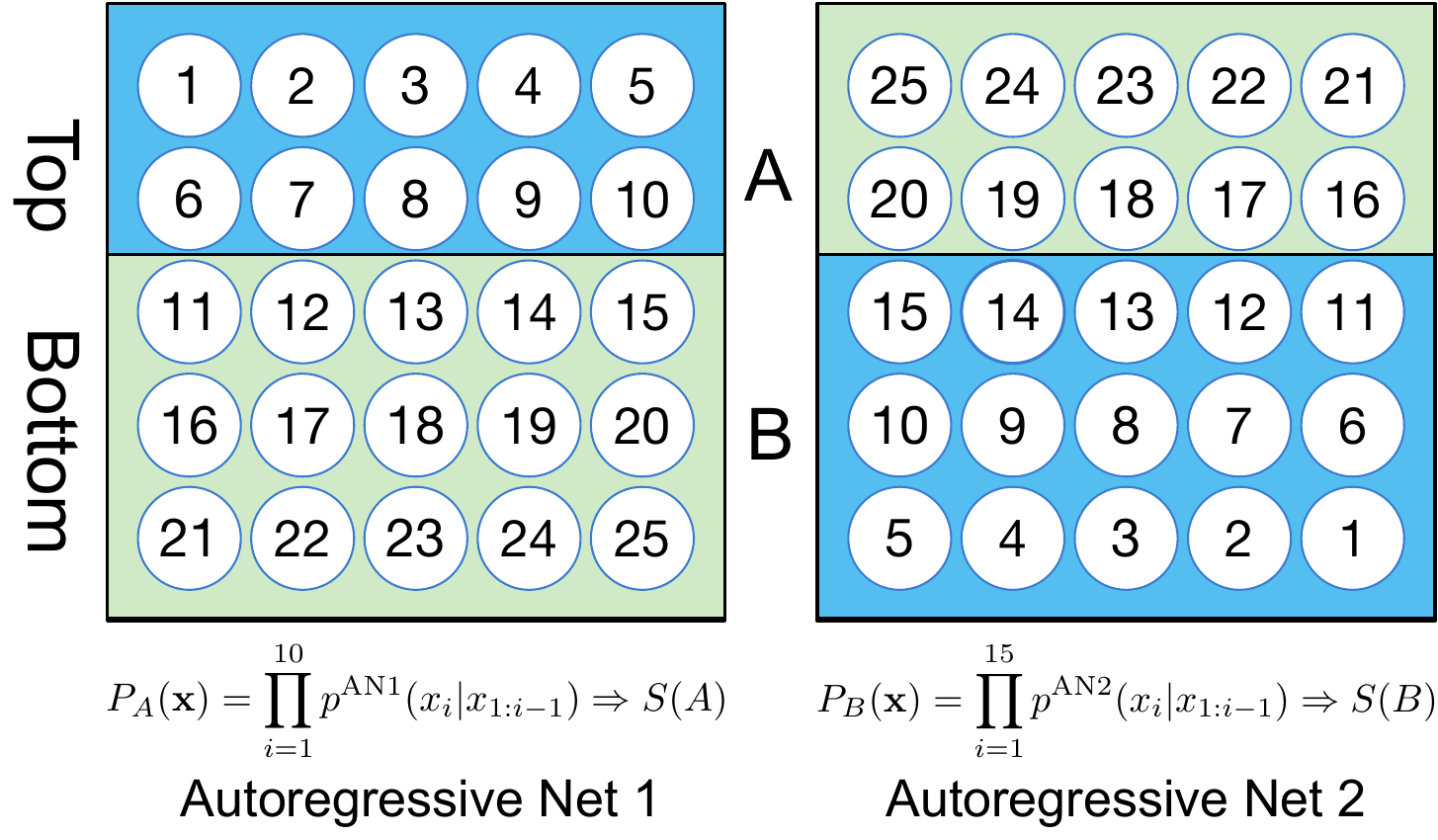}
    \caption{Autoregressive neural networks and two distinct orderings for computing marginal probabilities. Left: Autoregressive Network~1 processes a $5\times5$ image in raster scan order ($1$ to $25$), computing conditional probabilities $\mathbb{P}_A(\boldsymbol{x})=\prod_{i=1}^{10} p^{\text{AN1}}(x_i|x_{1:i-1})$ to estimate entropy $S(A)$ of the top region. Right: Network~2 uses reverse ordering ($25$ to $1$) to compute $\mathbb{P}_B(\boldsymbol{x})=\prod_{i=1}^{15} p^{\text{AN2}}(x_i|x_{1:i-1})$ for entropy $S(B)$ of the bottom region. These two networks with complementary orderings are trained separately, allowing estimation of marginal entropies for both top and bottom regions independently and thus mutual information via Eq.~\eqref{eq:entropy_MI_AN}.}
    \label{fig:MI:autoregressive}
\end{figure}
Training an autoregressive neural network involves maximizing the likelihood of the observed data by optimizing the parameters $\boldsymbol{\theta} = \{\theta_1, \theta_2, \ldots\}$:
\begin{equation}
    \arg \max_{\boldsymbol{\theta}\in\mathcal{M}} \frac{1}{|\mathcal{D}|}\sum_{\boldsymbol{x}\in \mathcal{D}}\log \mathbb{P}_{\boldsymbol{\theta}}(\boldsymbol{x})=\frac{1}{\vert \mathcal{D} \vert} \sum_{\boldsymbol{x} \in\mathcal{D} }\sum_{i=1}^n\log \mathbb{P}_{\theta_i}(x_i \vert \boldsymbol{x}_{< i}),
\end{equation}
where we have substituted the factorized joint distribution of an autoregressive model [Eq.~\eqref{eq:pARN}].

After training, we can compute entropies using the learned conditional probabilities. Given that all conditional probabilities are normalized, we can calculate entropies of subregions respecting the sequential ordering through Monte Carlo sampling. However, estimating MI remains challenging as the sequential structure prevents obtaining arbitrary marginal density functions. With a trained network, we can estimate $\log \mathbb{P}^{\text{AN1}}(x)$ and $\log \mathbb{P}^{\text{AN1}}(x,y)$. We then train another network with reverse ordering to access $\log \mathbb{P}^{\text{AN2}}(y)$ and $\ln \mathbb{P}^{\text{AN2}}(y,x)$ for points $x$ and $y$ in $A$ and $B$, respectively (Fig.~\ref{fig:MI:autoregressive}). The MI is then estimated as
\begin{equation}
    \begin{aligned}
        I(A:B) & = S^{\text{AN1}}(A) + S^{\text{AN2}}(B)                     \\
               & \quad -\frac{S^{\text{AN1}}(A,B) + S^{\text{AN2}}(A,B)}{2}.
    \end{aligned}
    \label{eq:entropy_MI_AN}
\end{equation}
For this estimator to be reliable, consistency between the two model distributions must be ensured. As shown in Fig.~\ref{fig:entropy:images}~(a), the difference in estimated entropies at $L = L_{\max}$ is negligible for MNIST, Fashion-MNIST, and CIFAR-10 datasets, indicating close distributions.

We employ PixelCNN~\cite{VanDenOord2016Conditional,Poolel2019Variational} and PixelCNN++~\cite{Salimans2017PixelCNN} architectures~\cite{pixel_models} for estimating conditional probabilities. These models process images in a top-to-bottom sequence but are not compatible with a spiral processing path needed for center-surrounding partitions. PixelCNN assumes a discrete data distribution, requiring discretization of pixel values into 256 bins, processed using a logistic mixture likelihood model. This discretization introduces a scaling factor in the MI estimates, absent in models like MINE and kNN that assume continuous distributions. We will adjust the results from PixelCNN and PixelCNN++ in Fig.~\ref{fig:MI:images} by a common scaling factor for consistency with MINE and kNN models.

\subsection{Estimation from Samples: Mutual Information Neural Estimation (MINE)}

MINE is a parametric estimator that employs variational neural networks to estimate MI~\cite{Belghazi2018MINE}. It is particularly effective for complex partitions like center:surroundings in images and text, and serves as a benchmark for autoregressive models in simpler partitions like top:bottom. The idea behind MINE is interpreting mutual information as the Kullback-Leibler (KL) divergence between joint and marginal distributions, transformed into a dual representation. Mutual information is represented as the KL divergence between the joint, $\mathbb{P}_{AB}$, and the product of the marginals, $\mathbb{P}_{A}\otimes\mathbb{P}_{B}$: $I(A:B) = D_{\text{KL}}({\mathbb{P}_{AB}}\mid\mid{\mathbb{P}_{A}\otimes \mathbb{P}_{B}})$, where $D_{\text{KL}}$ is defined as $D_{\text{KL}}({\mathbb{P}}\mid\mid{\mathbb{Q}}) := \mathbb{E}_{\mathbb{P}}\left[ \log \frac{\dd \mathbb{P}}{\dd \mathbb{Q}}\right]$. Applying the Donsker-Varadhan dual representation of the KL divergence~\cite{Donsker1983Asymptotic}:
\begin{equation}
    D_{\text{KL}}(\mathbb{P}\|\mathbb{Q}) = \sup_{T : \Omega \to \mathbb{R}} \left(\mathbb{E}_{\mathbb{P}}[T] - \log(\mathbb{E}_{\mathbb{Q}}[e^{T}])\right),
\end{equation}
where the supremum is taken over all functions $T : \Omega \to \mathbb{R}$ such that the two expectations are finite. Hence,
\begin{equation}
    I_{\text{MINE}}(A,B):=\sup_{\theta\in\Theta} \left(\mathbb{E}_{\mathbb{P}_{AB}}[T_{\theta}]-\log(\mathbb{E}_{\mathbb{P}_{A}\otimes \mathbb{P}_{B}}[e^{T_{\theta}}])\right).
    \label{eq:defMINE}
\end{equation}
By limiting the class of score functions $T_\theta$ to those represented by deep neural networks $\theta$, Eq.~\eqref{eq:defMINE} provides a lower bound on mutual information~\cite{Belghazi2018MINE}. This bound is tight for the optimal score function $T^*$, and enhancing the expressive power of neural networks ensures that MI can be approximated to the desired accuracy. As a lower-bound estimator, a higher MI value using a different network provides a better estimation, allowing for systematic improvement of results.

In practice, this lower bound is estimated on the entire dataset $\mathcal{D}$, with optimization solved by stochastic minibatch gradient descent. The optimized $I_{\text{MINE}}(A,B)$ is considered a lower bound of true MI.

MINE naturally aligns with two-category classification tasks, such as image or text classification, where the neural network outputs a real number indicating the classification result. In the original work of Ref.~\cite{Belghazi2018MINE}, a fully connected feedforward neural network was used to represent score functions. We incorporate convolutional neural networks as our score functions $T_{\theta}$, which have proven highly effective for image~\cite{Krizhevsky2012Imagenet} and text~\cite{Kim2014Convolutional} classification. This allows us to achieve improved variational estimates (see Appendix~\ref{App:MINE} for comparison) and scale up our calculations. The mutual information neural estimator is versatile and applicable to both top:bottom and center:surroundings partitions.

\subsection{Estimation from Samples: kNN}

The $k$-nearest neighbor (kNN) estimator is a nonparametric method that approximates the data distribution by assuming it is constant within high-dimensional simplices defined by the $k$ nearest neighbors. The density $\hat{\mathbb{P}}(x)$ at a point $x$ is estimated as:
\begin{equation}
    \hat{\mathbb{P}}(x)\approx \frac{k}{M\,\text{Vol}_{\text{kNN}}(x)},
    \label{eq:KNN}
\end{equation}
where $\text{Vol}_{\text{kNN}}(x)$ is the volume covering the $k$ nearest data points to $x$ out of $M$ samples. This estimated distribution is used to compute Shannon entropy and, consequently, the MI of the datasets. We use a refined kNN estimator~\cite{Kraskov2004Estimating,npeet_knn_estimator}. Due to implementation specifics, MI is determined up to an additive constant that depends on $k$ and the sample number $n_\text{data}$ (see Appendix~\ref{App:kNN}). We adjust our results globally to account for this when comparing with MINE and autoregressive estimators (see Fig.~\ref{fig:MI:images}).

The kNN estimator performs well on low-dimensional data but its accuracy diminishes with increasing data dimensionality. Nevertheless, it can provide MI estimates for text data and all image partitions considered in our study.

In the following sections, we utilize these estimators to analyze mutual information scaling in text and image datasets, assessing the viability of tensor network representations based on the observed scaling behaviors.

\section{Mutual Information Scaling in Images}
\label{sec:MI_image}

Images inherently possess spatial correlations reflecting real-world relationships. For example, in a face image, an eye on one side implies the presence of another on the opposite side. These correlations occur at various scales, with short-range correlations being more prevalent. Such short-range correlations are typically captured by the initial layers of convolutional neural networks~\cite{Krizhevsky2012Imagenet}. Consequently, we hypothesize that the mutual information in images will scale close to an area law, with possible additional contributions from longer-range correlations.

To test this hypothesis, we employ autoregressive network modeling, MINE, and kNN MI estimators to analyze low-resolution real-world image datasets. These include the MNIST handwritten digit dataset~\cite{mnist}, the Fashion-MNIST clothing images~\cite{Xiao2017Fashionmnist}, and the CIFAR-10 dataset~\cite{cifar}, which comprises natural images of animals and vehicles. These datasets are widely used to benchmark machine learning approaches.

\subsection{Entropy Scaling}

\begin{figure}[ht]
    \includegraphics[width=0.95\linewidth]{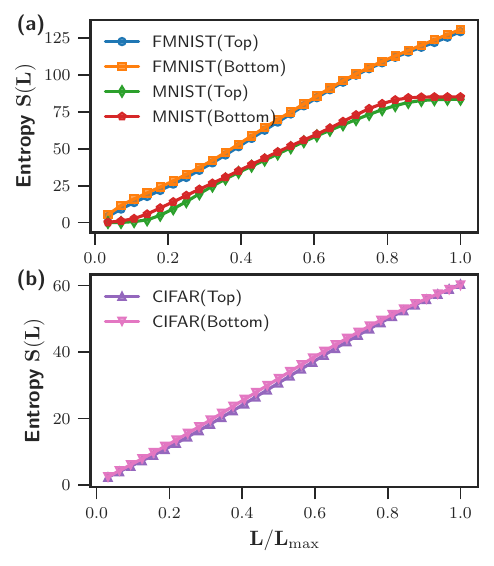}
    \caption{Shannon entropy scaling in image datasets estimated using PixelCNN~\cite{PixelRCNN} and PixelCNN++ architectures~\cite{Salimans2017PixelCNN}. (a) Entropy curves for MNIST and Fashion-MNIST ($28\times28$ pixels) showing volume law scaling $S(L) \propto L$ for both top and bottom regions. The close agreement between top and bottom curves indicates consistent probability distributions captured by the trained models. (b) Corresponding analysis for CIFAR-10 ($32\times32\times3$ pixels) demonstrating similar volume law scaling. The $x$-axis shows normalized region length $L/L_{\max}$, while the $y$-axis displays entropy $S(L)$ in bits.}
    \label{fig:entropy:images}
\end{figure}

We begin by examining the entropy of subregions within image data using trained autoregressive neural networks (elaborated in Sec.~\ref{sec:AutoregressiveNetworks}). The results are presented in Fig.~\ref{fig:entropy:images} for the top-bottom ($\text{T}:\text{B}$) partitioning of MNIST, Fashion-MNIST, and CIFAR-10 datasets. We include the results obtained from autoregressive networks of both orderings.

From Fig.~\ref{fig:entropy:images}, we observe that entropy scales similarly to thermal entropy in physical systems, adhering to a volume law. The volume law in entropy supports our hypothesis that MI is a more suitable metric for studying the information structure in classical data. The closeness of the entropy curves for the two orderings also supports the consistency of the probability distributions captured by the trained autoregressive models.

Image datasets typically focus on a central object with fewer distinctive features towards the edges. This effect is particularly pronounced in MNIST, where edges are often blank. This is evident in Fig.~\ref{fig:entropy:images}(a), which shows a minimal slope near the boundaries ($L\approx 0$ or $L\approx L_{\max}$). As discussed in Appendix~\ref{App:kNN}, the MI of MNIST data also decreases towards the edges due to the lack of features. To mitigate this edge effect, we analyzed the data after making the images translationally invariant by randomly displacing the images, as depicted in Fig.~\ref{fig:A-schematic-illustration}. We note that this effect is less pronounced in datasets with natural images.

\subsection{Mutual Information Scaling}

\begin{figure}[ht]
    \includegraphics[width=0.99\linewidth]{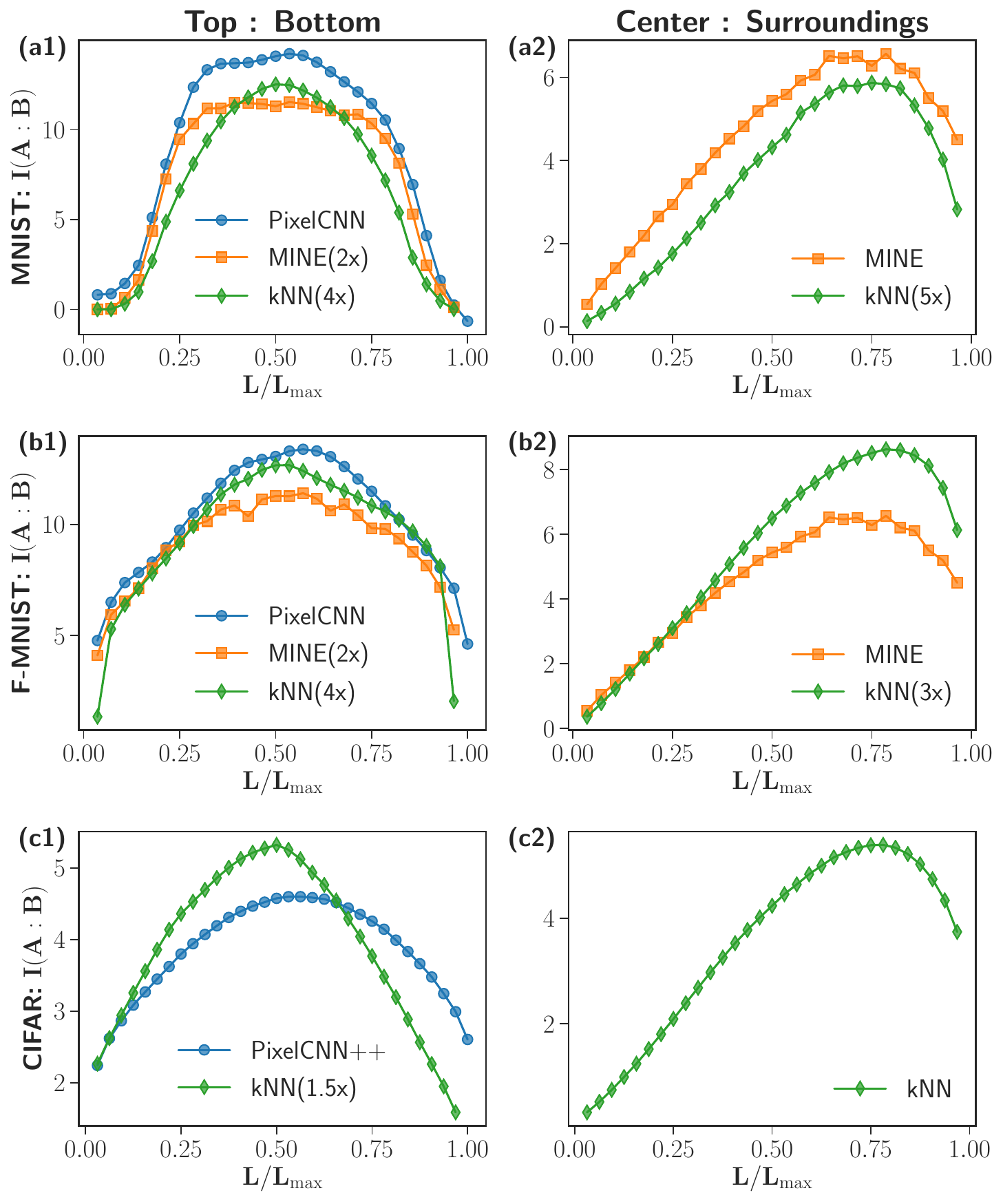}
    \caption{Mutual information scaling in image datasets analyzed using three complementary estimation methods: (i) PixelCNN/PixelCNN++ autoregressive networks computing exact conditional probabilities, (ii) mutual information neural estimation (MINE) using convolutional neural networks as variational functions, and (iii) $k$-nearest neighbor (kNN) density estimation. Left panels show top:bottom ($\text{T}:\text{B}$) partitioning; right panels show center-surroundings ($\text{C}:\text{S}$) partitioning for: (a1--a2)~MNIST, (b1--b2)~Fashion-MNIST ($28\times28$ pixels), and (c1--c2)~CIFAR-10 ($32\times32\times3$ pixels). For MNIST, the simplest dataset, we observe evidence of area law scaling through $\text{T}:\text{B}$ saturation and linear $\text{C}:\text{S}$ growth. For more complex datasets (Fashion-MNIST and CIFAR-10), $I(\text{C}:\text{S})$ still grows linearly but $I(\text{T}:\text{B})$ does not reach a plateau, suggesting faster than area law scaling. The $x$-axis represents the normalized partition length $L/L_{\max}$, and the $y$-axis shows the mutual information $I(A:B)$ in bits, with kNN results globally adjusted for consistent comparison.}
    \label{fig:MI:images}
\end{figure}

Figure~\ref{fig:MI:images} presents the mutual information (MI) curves for the top-bottom ($\text{T}:\text{B}$) and center-surroundings ($\text{C}:\text{S}$) partitions on the translationally invariant MNIST dataset. The $I(\text{T}:\text{B})$ curves display a noticeable plateau in the central region, indicative of area law scaling. Concurrently, the $I(\text{C}:\text{S})$ curves grow linearly at smaller $L$ values, further supporting the area law hypothesis.

We extend our investigation to more complex image datasets, including Fashion-MNIST and CIFAR-10. Similar to MNIST, we randomly displaced the images to ensure translation invariance. We employed MINE and PixelCNN++ autoregressive networks, along with kNN methods for Fashion-MNIST. The CIFAR-10 dataset, with its three color channels, increases the data dimensionality, presenting an additional challenge for MI estimation. Although MINE results for CIFAR-10 are not available due to computational constraints, our analysis indicates that while $I(\text{C}:\text{S})$ continues to grow linearly, $I(\text{T}:\text{B})$ does not reach a plateau (Fig.~\ref{fig:MI:images}). This observation suggests that in more generic images, MI scaling could exceed the area law.

These findings have important implications for tensor network representations of image data. The area law scaling observed in MNIST suggests that two-dimensional tensor networks like PEPS could efficiently represent this dataset, aligning with previous successful applications of tensor network algorithms to MNIST classification tasks~\cite{Stoudenmire2016Supervised,Huang2017Provably,Glasser2019Expressive,Cheng2021Supervised} and attempts in modeling MNIST images~\cite{Han2018Unsupervised,Vieijra2022Generative}. However, the faster-than-area-law scaling in more complex datasets like Fashion-MNIST and CIFAR-10 indicates that traditional tensor network approaches may face scalability challenges for larger, more complex images. This finding motivates exploring more generalized tensor network states or hybrid models that combine tensor networks with neural networks~\cite{Liu2023Tensor}, potentially offering a solution to these scalability issues while maintaining computational efficiency.

\section{Mutual Information Scaling in Text}
\label{sec:MI_text}

Natural text is inherently complex due to factors like grammatical structure, semantics, style, and cross-references, leading to correlations at various scales--from sentence-level grammar to paragraph or document-level coherence. These multi-scale correlations suggest high mutual information between different text segments. Supporting this, recent studies have observed algebraically decaying correlations at both the character and word levels~\cite{Lin2017Critical,Lin2017Why,Shen2019Mutual}. Advanced transformer models~\cite{Vaswani2017Attention} capable of generating long, coherent text also exhibit power-law mutual information scaling. In contrast, models that struggle with long coherence, such as recurrent neural networks~\cite{Shen2019Mutual} and long short-term memory networks (LSTM)~\cite{LSTM}, show exponentially decaying correlations.

To accommodate the distinct structure of text data, we utilize two MI estimation methods introduced in Sec.~\ref{sec:MI_est}: the mutual information neural estimator (MINE)~\cite{Belghazi2018MINE} and the $k$-nearest neighbor (kNN) estimator~\cite{Kraskov2004Estimating}. We apply these methods to a dataset consisting of Wikipedia articles and analyze the mutual information scaling. Subsequently, we introduce a random pair model and a dependency tree model as toy models of text data to understand the observed scaling.

\subsection{Power Law Scaling in Text}
\label{sec:MI_classical_text}

We analyze the WikiText-2 dataset, which consists of 600 training articles and 2 million tokens~\cite{Merity2016Pointer}. Words are converted into a computer-readable format using pre-trained word-level embeddings from the 50-dimensional GloVe model~\cite{Pennington2014Glove}. This model generates dense vectors for each word, ensuring that words appearing in similar contexts, and thus sharing similar meanings, are proximate in the feature space. Consequently, our mutual information estimation incorporates word meanings. We have verified the robustness of our results by testing them against changes in the dimensionality of the embedding space to $200$.

We employ the kNN and MINE estimators, using both fully connected and convolutional neural networks as variational functions in the estimator, as detailed in Sec.~\ref{sec:MI_est}. Both methods yield nearly identical results. Initially, we utilized a fully connected feedforward neural network as the score function, in line with the original version of Ref.~\cite{Belghazi2018MINE}. To stabilize the optimization, we applied the moving average gradient trick mentioned in Refs.~\cite{Belghazi2018MINE,Poolel2019Variational}. Subsequently, we improved the mutual information neural estimator by using a text convolutional neural network~\cite{Kim2014Convolutional} as the score function. While this introduces some biases into the estimation, the results estimated with CNNs are larger and thus more accurate than those estimated with feedforward neural networks. Additionally, CNNs have fewer tunable parameters than feedforward networks, enabling us to scale up to deeper and wider networks.

\begin{figure}
    \includegraphics[width=0.99\linewidth]{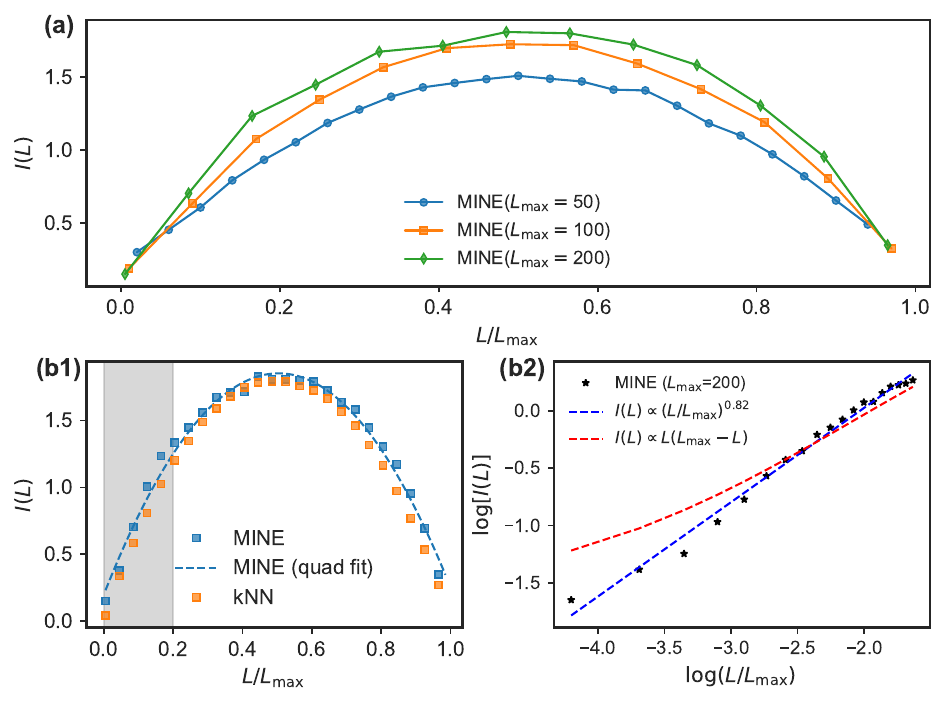}
    \caption{Mutual information analysis of the WikiText-2 dataset using 50-dimensional GloVe word embeddings. (a) MINE estimates for varying sequence lengths $L_{\max}=50,100,200$ words, showing consistent scaling behavior after normalization. A text convolutional neural network~\cite{Kim2014Convolutional} is used as the score function. (b1) Comparison between MINE and kNN ($k=20$, $\text{MaxNText}=10000$) estimators for $L_{\max}=200$, with power-law fitting region highlighted in gray. (b2) Log-log analysis of the initial part of the $I(L)$ curve revealing power-law scaling $I(L)\propto L^{0.82(2)}$ (blue dashed line) for small $L$, compared with the theoretical upper bound $I(L)\propto L(L_{\max}-L)$ (red dashed line) derived for maximally correlated elements in Sec.~\ref{Sec:ToyModel}. The $x$-axis represents the normalized length of the left region, $L/L_{\max}$, and the $y$-axis shows the mutual information $I(\text{L}:\text{R})$ in bits.}
    \label{fig:MI:Wikitext}
\end{figure}

As anticipated, the strong correlations between different segments of the text result in an MI scaling that is significantly steeper than the logarithmic scaling observed in critical systems, as shown in Fig.~\ref{fig:MI:Wikitext}~(b)--(c). Specifically, for the WikiText-2 dataset, we observe power-law correlations for small lengths $L$, with an exponent $\nu = 0.82(2)$. This scaling is nearly equivalent to a volume law, where MI grows linearly with the system size. For context, an area law scaling would result in a constant MI, independent of system size. Furthermore, as illustrated by the dashed line in Fig.~\ref{fig:MI:Wikitext}~(c), the scaling closely aligns with a model where all words are equally correlated, regardless of their distance. This would result in a scaling of $I(\text{L}:\text{R})\propto L\,(L_{\max}-L)$.

These findings have significant implications for representing text using tensor networks. The observed power-law scaling, approaching a volume law, suggests that traditional one-dimensional tensor network approaches, such as matrix product states (MPS) or tree tensor networks (TTN), may not scale efficiently to long texts. This is because these tensor network structures are designed to capture area law or logarithmic scaling of entanglement, which is much slower than the observed near-volume-law scaling in text data.

Moreover, our results challenge the assertion made by Ref.~\cite{Lin2017Critical} that languages exhibit critical distributions, implying that the structures of natural languages significantly differ from local critical systems. In quantum critical systems, power-law decaying correlations typically lead to logarithmic scaling of entanglement entropy~\cite{Calabrese2004Entanglement,CalabreseCardy2009,CasiniHuerta2009}. However, our observations show that in classical text data, power-law correlations coexist with near-volume-law scaling of mutual information. This discrepancy highlights the fundamental differences between quantum and classical systems in terms of information structure.

Another significant observation from our findings is the universal scaling of the entire distribution, as shown in Fig.~\ref{fig:MI:Wikitext}~(b). The consistency of the MI curve as we increase $L_{\max}$, with only a constant rescaling of the overall height, suggests that high levels of mutual information persist even in longer texts. For instance, doubling the text length from $L_{\max} = 100$ to $L_{\max} = 200$ results in a similar MI curve shape, merely scaled by a constant factor. This scaling behavior indicates that the information structure of text remains consistent across different length scales.

However, we must also consider the potential existence of an intermediate characteristic scale $L_{\max}$, beyond the scope of our numerical analysis, where correlations might exhibit a different decay pattern. Indeed, previous research by \citet{Shen2019Mutual} identified a length scale at which correlations between individual words disappear. Therefore, it is conceivable that significantly larger values of $L_{\max}$ could reveal a different scaling behavior. This possibility underscores the need for further investigation into the information structure of text at various scales.

\subsection{Scaling of Correlation Functions and a Random Pair Model for Text}
\label{Sec:ToyModel}

The observed power-law scaling of MI in text data, combined with the known algebraic decay of correlations between characters or words from the literature~\cite{Lin2017Critical,Shen2019Mutual}, presents an intriguing puzzle. This behavior differs significantly from what is typically observed in quantum critical systems, where power-law correlations usually lead to logarithmic scaling of entanglement entropy. To better understand this phenomenon and its implications for modeling natural language, we first review the role of correlation functions in quantum and classical systems, then introduce a simplified toy model that captures these key features of text data.

In principle, any probability distribution can be decomposed into the form of Eq.~\eqref{eq:pARN} by tabulating every conditional probability $\mathbb{P}(x_i \vert \boldsymbol{x}_{<i})$. However, this approach becomes inefficient for large systems due to the exponential growth in parameters. Early language models, such as $n$-grams, addressed this by limiting connectivity:
\begin{equation}
    \mathbb{P}\left(x_{t+1} \mid x_{t}, \ldots, x_{1}\right)=\mathbb{P}\left(x_{t+1} \mid x_{t}, \ldots, x_{t-n+2}\right).
    \label{eq:ngram}
\end{equation}
This Markovian property allows for matrix product state representations~\cite{Kato2019Quantum,Glasser2019Expressive,Gao2022Enhancing}, resulting in area law scaling of mutual information and exponential decay of correlations.

Matrix product states (MPS) naturally reproduce the typical decay of correlations characteristic of gapped systems, which further explains why MPS effectively represent ground states of gapped models. Specifically, the correlations between two sites $i$ and $j$ are primarily created through the tensors in the shortest path connecting them. Mathematically, via transfer operators, the two-point correlation function $C_{\text{MPS}}(A_i,B_j)=\langle A_{i}B_{j} \rangle - \langle A_{i} \rangle \langle B_{j} \rangle$ in a constant bond dimension MPS decays exponentially in the asymptotic limit:
\begin{equation}
    C_{\text{MPS}}(A_i,B_j)\propto e^{-|i-j|/\xi},
\end{equation}
for some correlation length $\xi>0$.

Previous research~\cite{Gallego2022Language,Lin2017Why,Lin2017Critical,Shen2019Mutual} has identified algebraic decay in correlations between characters or individual words in natural language, a feature reminiscent of critical systems in physics~\cite{CardyBook1996}. In these critical physical systems, gapless excitations with infinite range lead to power-law decaying correlations
\begin{equation}
    C(A_i,B_j)\propto |i-j|^{-\alpha},
\end{equation}
where $\alpha\geq 0$ is some exponent. In quantum critical models, this implies logarithmic scaling of the entanglement entropy~\cite{Calabrese2004Entanglement,CalabreseCardy2009,CasiniHuerta2009}. For a finite system with $N$ sites, it is possible to increase the bond dimension $D$ of the matrix product state polynomially with $N$ to reproduce algebraic correlations at long distances. In contrast, in tree tensor networks and multi-scale entanglement renormalization ansatz (MERA), correlations decay algebraically, as required in gapless models.

This observation led Lin and Tegmark to construct a character-level statistical language model that exhibits similar long-range correlations~\cite{Lin2017Why,Lin2017Critical}. In their model, long-range correlations emerge from hidden variables representing linguistic structures and meanings. This binary Markov tree-based model can then be represented by tree tensor networks (TTNs), and is expected to result in power-law correlations and logarithmic mutual information scaling. Based on this analogy, it has been argued that natural languages exhibit critical behavior~\cite{Gallego2022Language,Lin2017Critical}, and therefore one could expect languages to behave like a critical quantum system. Therefore, languages should possess critical properties of mutual information, which can be considered analogous to entanglement in classical systems~\cite{Wilms2011Mutual}.

However, our observations indicate a much steeper growth of mutual information, suggesting that long-range correlations play a more significant role than assumed by Lin and Tegmark. This might seem surprising at first, given that critical physical systems typically exhibit logarithmic scaling of entanglement entropy~\cite{Calabrese2004Entanglement,CalabreseCardy2009,CasiniHuerta2009}. We argue that the algebraic scaling of mutual information in classical data does not contradict the presence of algebraic correlations. This difference arises because natural language data does not exhibit the notion of locality typically imposed in quantum systems, leading to a broader range of possible mutual information scaling behaviors.

To address this puzzle, we introduce a random pair model, a classical analogue of the random singlet model introduced in Ref.~\cite{Wolf2008Area}. This model demonstrates how algebraic correlations in classical probability distributions can coexist with power-law mutual information scaling, in contrast to the logarithmic scaling typically observed in quantum critical systems. This is an initial minimal model reproducing our observations but does not take linguistic structure into account. We capture those aspects in a more realistic model in Sec.~\ref{Sec:DepTreeModel}.

\begin{figure}[ht]
    \centering
    \includegraphics[width=0.9\linewidth]{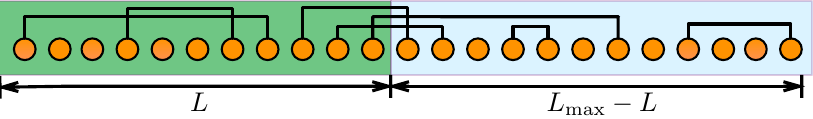}
    \caption{A schematic presentation of the random pair model. Lattice sites form maximally correlated pairs with one randomly chosen other site following a probability distribution $\rho(x-y)$. The mutual information $I(A:B)$ between two regions $A$ and $B$ is given by the number of links connecting them.}
    \label{fig:MI:ToyModel}
\end{figure}

Consider a probability distribution describing a one-dimensional lattice of classical degrees of freedom, such as words. Each word is assigned a coordinate $x$. The probability distribution is characterized by each site $x$ being correlated with only one other site $y$, with the pair $(x,y)$ sharing the maximum possible mutual information (Fig.~\ref{fig:MI:ToyModel} provides an illustration). The correlated pairs are randomly distributed across the lattice, following a probability distribution $\rho(x-y)=C|x-y|^{-\alpha}$, where $\alpha>1$ is a model parameter, and $C$ is a normalization constant. This model exhibits algebraically decaying correlation functions $\propto|x-y|^{-\alpha}$.

In this model, the mutual information is determined by the number of pairs where one element is in region L and the other in R. Hence, we can express the mutual information as:
\begin{equation}
    I(\text{L}:\text{R}) = \sum_{x=1}^{L}\sum_{y=L+1}^{L_{\max}}\rho(x-y).
    \label{eq:MI_sum}
\end{equation}
where $L_{\max}$ is the total length of the system. To extract the leading asymptotic behavior, we can approximate the sum over $y$ by an integral:
\begin{equation}
    \rho_L(z) \equiv \sum_{w=z}^{\infty}\rho(w) \approx \int_z^{\infty} C w^{-\alpha} dw = \frac{C}{\alpha-1} z^{1-\alpha},
    \label{eq:rho_L}
\end{equation}
for $z > 0$. This allows us to simplify the expression for mutual information:
\begin{equation}
    I(\text{L}:\text{R}) \approx \sum_{x=1}^{L} \rho_L(L-x+1) = \frac{C}{\alpha-1} \sum_{x=1}^{L} (L-x+1)^{1-\alpha}.
    \label{eq:MI_simplified}
\end{equation}
For large $L$, we can approximate this sum by an integral:
\begin{equation}
    I(\text{L}:\text{R}) \approx \frac{C}{\alpha-1} \int_0^L (L-x)^{1-\alpha} \dd x = \frac{C}{(\alpha-1)(2-\alpha)} L^{2-\alpha}.
    \label{eq:MI_final}
\end{equation}
This result shows that the model exhibits both algebraic correlations and algebraic scaling of mutual information. The exponent of the mutual information scaling, $2-\alpha$, is directly related to the exponent of the correlation decay, $\alpha$. This relationship provides insight into how different correlation structures in the data can lead to various mutual information scaling behaviors: (i) For $1 < \alpha < 2$, we observe a power-law scaling of mutual information with an exponent between 0 and 1. This regime corresponds to our observations in the WikiText-2 dataset, where we found $\nu \approx 0.82$, implying $\alpha \approx 1.18$; (ii) For $\alpha = 2$, the model would result in logarithmic mutual information scaling, reminiscent of critical quantum systems; (iii) For $\alpha > 2$, the mutual information would saturate to a constant value for large $L$, corresponding to an area law.

The flexibility of this model in producing different scaling behaviors highlights the rich structure possible in classical data, which can differ significantly from quantum systems. Note that while this random pair model captures the observed mutual information scaling, it may not have an efficient matrix product state representation. However, it can be efficiently represented by restricted Boltzmann machines~\cite{Gao2017Efficient}, suggesting that alternative network architectures might be more suitable for capturing the information structure of natural language.

An interesting limiting case of the random pair model occurs when pairs are uniformly distributed, i.e., $\alpha=0$. In this scenario, the number of correlated pairs that a fraction of the system can form is proportional to the volume (length in 1D) of that fraction. Consequently, the number of correlated pairs between two fractions of the system is given by the product of their volumes. This is achieved in the random pair model by setting $\rho(x-y)=C$. As a result, the mutual information scales as:
\begin{equation}
    I(\text{L}:\text{R})=|\text{L}||\text{R}|\propto L(L_{\max}-L),
\end{equation}
where $L_{\max}$ is the total number of lattice sites. This curve serves as a benchmark for assessing the deviation of the data's probability distribution from a scenario where all elements are correlated.

\subsection{Dependency Tree Model}
\label{Sec:DepTreeModel}

While the random pair model provides a minimal framework for understanding MI scaling in text, it falls short in capturing the complex linguistic structures inherent in natural language. To provide a more realistic description, we introduce the \emph{dependency tree model}, a generative model that better reflects the grammatical and semantic relationships between words in a sentence and allows us to infer the length distribution of word-to-word mutual information (MI).

\begin{figure}[ht]
    \centering
    \includegraphics[width=0.6\linewidth]{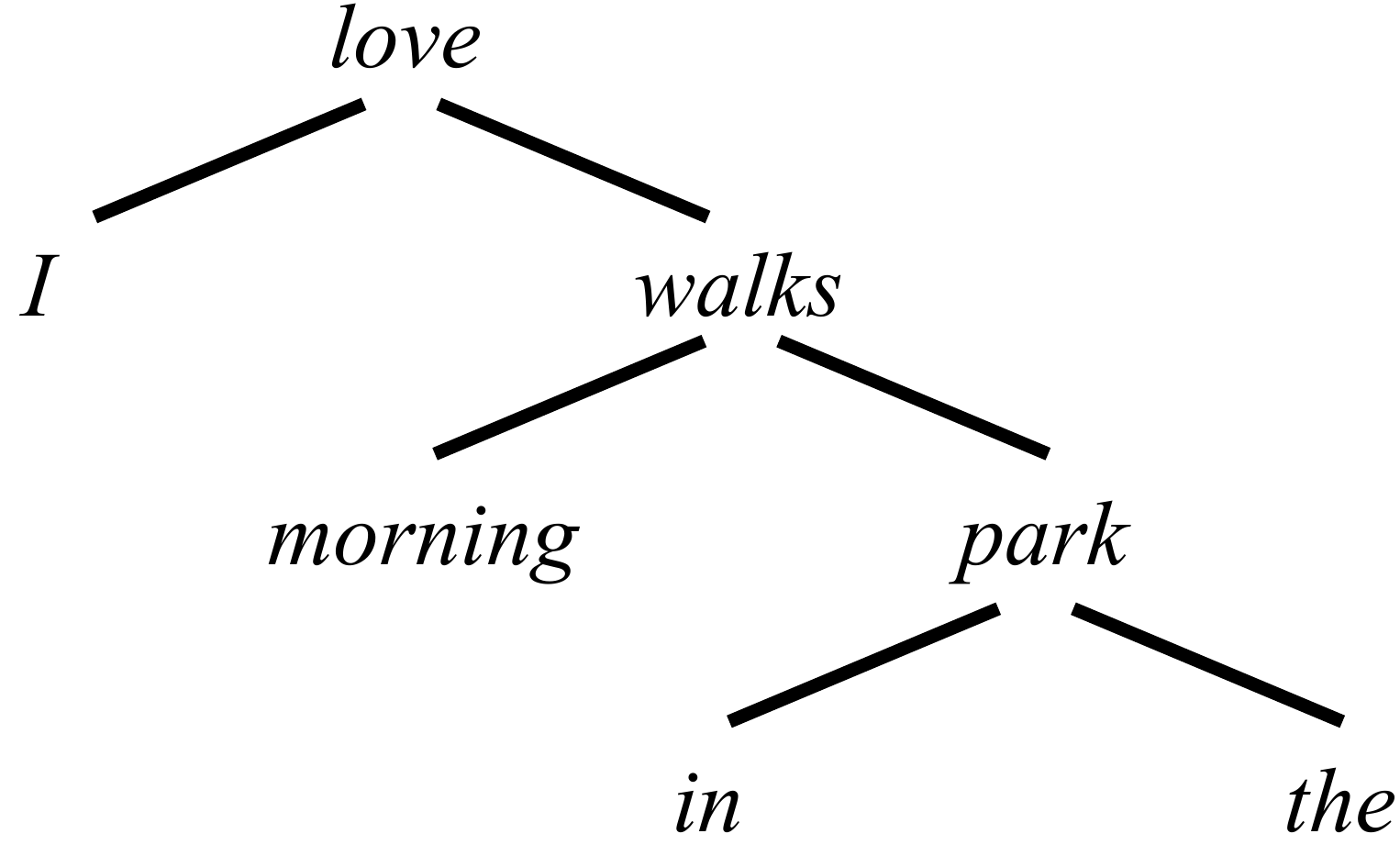}
    \caption{Example of a dependency parsing tree for the sentence ``I love morning walks in the park.'' The directed edges represent grammatical dependencies between words, forming a hierarchical structure. This tree representation motivates our dependency tree model (Sec.~\ref{Sec:DepTreeModel}), where both tree structure and word choices are generated probabilistically to capture linguistic patterns.}
    \label{fig:Dependency_parsing_example}
\end{figure}

In natural text, meaning and grammatical structure lead to complex correlations among words. Building on Chomsky's foundational work on formal grammars~\cite{Chomsky1957Syntactic,Chomsky1965Aspects}, a paradigmatic tool to capture these grammatical relations is \emph{dependency parsing}~\cite{Kubler2009Dependency,Nivre2005Dependency}. In dependency parsing, the syntactic structure of a sentence is modeled as a tree, where nodes represent words and edges denote directed grammatical relations between them. An example is presented in Fig.~\ref{fig:Dependency_parsing_example}.

\begin{figure}[ht]
    \includegraphics[width=0.8\linewidth]{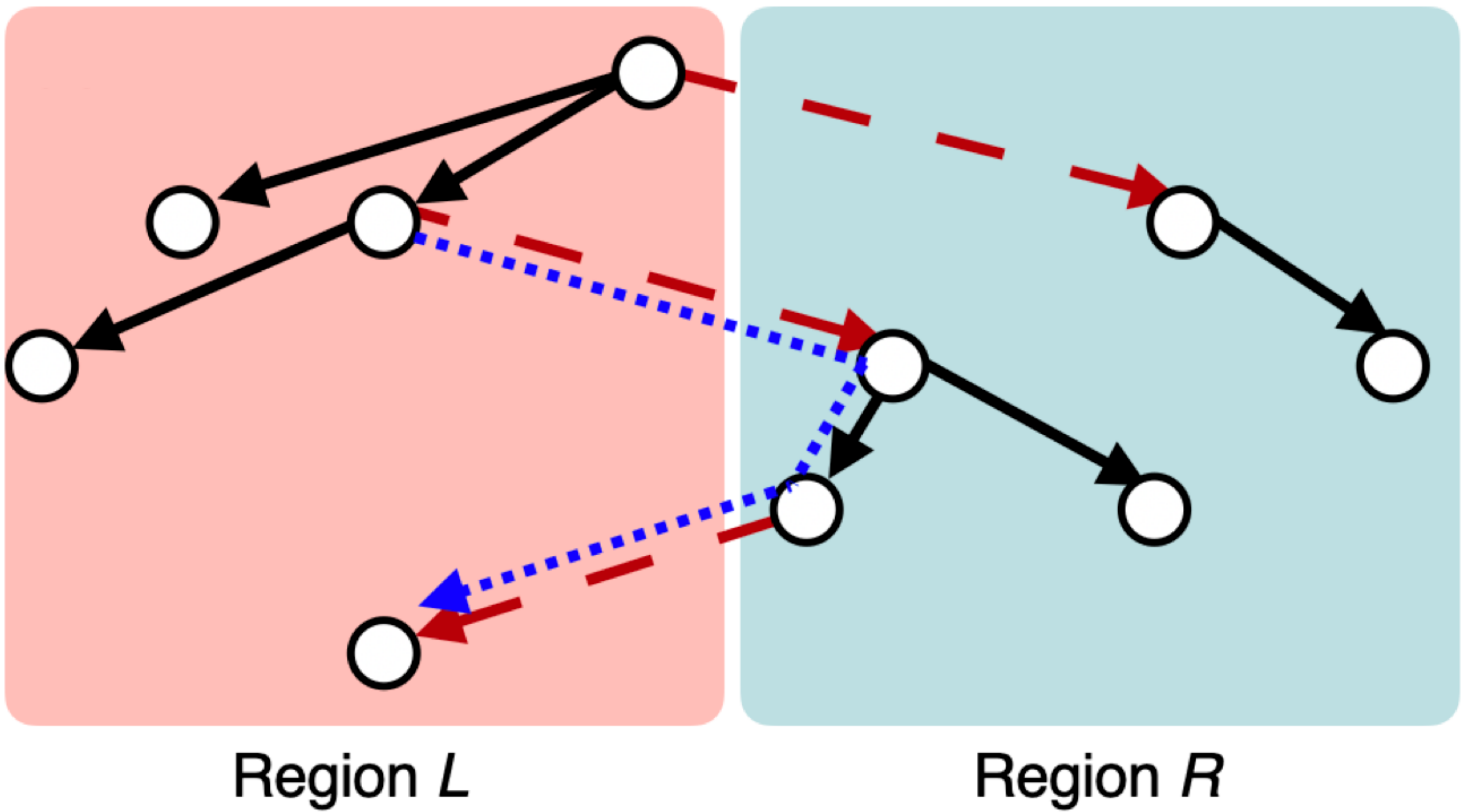}
    \caption{Schematic of the dependency tree model for mutual information calculation in text. Words (white circles) form nodes in a dependency parsing tree, with black arrows indicating grammatical dependencies sampled from a random distribution. The tree structure is sampled from a random distribution, while words are generated via a Markov process. For calculating mutual information between regions $L$ and $R$ via Eq.~\eqref{eq:MI_crossings_returns}, dashed red arrows indicate boundary crossings contributing positive MI terms $I(\text{crossing})$, while dotted blue lines represent return paths requiring subtraction of $I(\text{return})$ terms to avoid overcounting. Under the Markov assumption, these contributions fully determine the mutual information between regions.}
    \label{fig:Dependency_parsing_sec}
\end{figure}

Based on this grammatical structure, we introduce the \emph{dependency tree model}, a generative model for grammatical relations between words during text generation (see Fig.~\ref{fig:Dependency_parsing_sec}). In this model, the dependency tree structure is sampled from a random distribution (discussed below), and the words of the sentence are generated using a Markov model.

While the random pair model introduces correlations between pairs of words independently, the dependency tree model captures more complex linguistic dependencies that more closely resemble correlations in natural language. Within this model, MI scaling provides additional information about linguistic patterns. The discussion in this section up to Eq.~\eqref{eq:MI_crossings_returns} is general. By making some further assumptions, we analytically find that the length distribution of edges in the dependency parsing tree also follows a power law with an exponent $\nu - 2$, where $\nu$ is the exponent of the MI scaling. These simplifying assumptions are not mandatory; the model can be made more complex to capture additional aspects of natural text. In such cases, the MI scaling can be modeled numerically. The main results of this calculation are presented here, while detailed derivations can be found in Appendix~\ref{App:DepTreeModel}.

While earlier work by Lin and Tegmark~\cite{Lin2017Why,Lin2017Critical} proposed a generative model relying on a fixed, binary tree-shaped graphical model, our dependency tree model offers more flexibility. In their model, the bottom row of the graph represents words in a sentence, while higher nodes encode meaning and grammatical relations between the words. While this structure correctly predicts power-law correlations between individual words, it produces logarithmic mutual information scaling between regions, as the distributions can be expressed as tree tensor networks (TTNs), which contradicts our observations.

\subsubsection{Model Description and Assumptions}

The dependency tree model offers more flexibility than the binary tree-based structure of Refs.~\onlinecite{Lin2017Why,Lin2017Critical}, as it allows both the words and the tree structure to be sampled randomly. Each node in the tree corresponds to a word, and the structure of the tree represents the grammatical dependencies between words. We assume the Markov property within the graph; thus, each word is generated by its parent irrespective of nodes higher in the graph:
\begin{equation}
    \mathbb{P}\left(\text{child} \, \mid \, \text{all ancestor words}\right) = \mathbb{P}\left(\text{child} \, \mid \, \text{parent}\right).
\end{equation}
This simplification allows us to focus on how the randomly generated graph structure affects mutual information scaling. While this is a strong simplifying assumption, the model can be extended by allowing the nodes of the tree to carry additional information besides the word itself, such as meaning, similar to how recurrent neural networks use hidden state vectors to carry the meaning of a sentence.

\subsubsection{Mutual Information Contributions}

The Markov property significantly simplifies the calculation of the mutual information between two regions of text, $L$ (left) and $R$ (right). Specifically, there are only two types of contributions to the mutual information, which we call \emph{crossings} and \emph{returns}:
\begin{equation}
    I(L:R) = \sum_{\text{crossings}} I(\text{crossing}) - \sum_{\text{returns}} I(\text{return}),
    \label{eq:MI_crossings_returns}
\end{equation}
as we derive in Appendix~\ref{App:DepTreeModel}. These contributions are illustrated in Fig.~\ref{fig:Dependency_parsing_sec} as dashed and dotted arrows, respectively.

More formally, $I(\text{crossing}) = I(W_p : W_c)$ represents the mutual information between the parent word $W_p$ and the child word $W_c$ of an arrow crossing the boundary between $L$ and $R$. The correlation between these words introduces mutual information between the two regions, contributing positively to $I(L:R)$. However, these contributions are not independent, and simply summing them would lead to overcounting. To counter this, contributions from \emph{returns} must be subtracted. \emph{Returns} correspond to the shortest path connecting a parent word $W_p$ at the boundary, before the path crosses to the other region, and its earliest descendant $W_d$ in the same region as the parent when the path returns. $I(\text{return})$ indicates the mutual information between the parent and the descendant word, $I(W_p : W_d)$. These MI contributions need to be subtracted to avoid overcounting, as this is the amount of information that is returned to the region where the path started. The derivation of Eq.~\eqref{eq:MI_crossings_returns} can be found in Appendix~\ref{App:DepTreeModel}.

\subsubsection{Mutual Information Scaling Between Regions}

To gain analytical insights into the mutual information scaling, we now introduce some simplifying assumptions. These assumptions, while not necessary for numerical simulations, allow us to derive analytical expressions that provide intuition about the model's behavior. The scaling in our model is determined by the random distribution generating the dependency tree. By comparing these analytical results with the power-law scaling of the mutual information observed in Sec.~\ref{sec:MI_classical_text}, we can infer properties of the dependency structure in real text.

To preserve the tree structure, each word in the generated graph can have at most one incoming and arbitrarily many outgoing edges. Let $q(L)$ be the probability distribution of the length $L$ of these edges, where $L$ is the length along the text direction (positive or negative depending on the direction of the edge). We neglect boundary effects and assume that $q(L)$ is uniform across the text.

Based on our discussion in Appendix~\ref{App:DepTreeModel} above Eq.~\eqref{eq:crossing_constant_MI}, we assume that all MI contributions from \emph{crossings} are identical. Furthermore, Eq.~\eqref{eq:return_MI_scaling} in Appendix~\ref{App:DepTreeModel} shows that under certain simplifying assumptions, the \emph{return} terms decay exponentially. Motivated by this, we neglect contributions from \emph{returns}. This approximation also serves as a strict upper bound, as these terms only decrease the mutual information in Eq.~\eqref{eq:MI_crossings_returns}.

Under these assumptions, the mutual information between two regions becomes proportional to the number of \emph{crossings} across the boundary. While these simplifications do not capture word-word correlation quantified by MI in real text with perfect accuracy, they provide a reasonable approximation for our analysis.

Let $\text{Cr}(L)$ be the average number of crossings between regions $L$ and $R$ of lengths $L$ and $L_{\max}-L$, where $L_{\max}$ is the total text length. Then:
\begin{align}
    \text{Cr}(L) &= \sum_{j=1}^{L_{\max}-1} \min(L, j, L_{\max}-j) \, q(j)\nonumber\\
                 &+ \sum_{j=1}^{L_{\max}-1} \min(L_{\max}-L, j, L_{\max}-j) \, q(-j).
\end{align}
where the first line corresponds to the expected number of crossings from $L$ to $R$ and the second line vice versa. Although this is not a simple expression to analyze, its first and second discrete derivatives are significantly more tractable. Assuming without loss of generality that the left region is shorter, $L < L_{\max}/2$, then the first discrete derivative is
\begin{equation}
    \text{Cr}(L+1) - \text{Cr}(L) = \sum_{j=L+1}^{L_{\max}/2} q(j) - \sum_{j=N/2+1}^{L_{\max}-L-1} q(-j).
\end{equation}
The second discrete derivative is particularly informative:
\begin{equation}
    \text{Cr}(L+2) + \text{Cr}(L) - 2\, \text{Cr}(L+1) = q(L+1) + q(L_{\max}-L).
    \label{eq:crossings_2nd_derivative}
\end{equation}
This is precisely the symmetrized edge length distribution. In our simple model, this must be proportional to the second derivative of the mutual information curve in Fig.~\ref{fig:MI:Wikitext}~(c). Based on the power-law scaling observed in Sec.~\ref{sec:MI_classical_text} for short and intermediate scales, we expect
\begin{equation*}
    q(L+1) + q(L_{\max}-L) \propto L^{\nu - 2}
\end{equation*}
for small $L$, where $\nu \approx 0.82$ as observed in the WikiText-2 dataset. We note, however, that our approximation of vanishing \emph{return} contributions is expected to be more accurate on longer scales.

Finally, we mention that the simplifying assumptions about the \emph{crossing} and \emph{return} terms were only necessary for the analytical calculation of the scaling. To model language more accurately, these assumptions can be relaxed. In such cases, numerical simulations based on Eq.~\eqref{eq:MI_crossings_returns} can provide more precise estimations of the scaling across all length scales, albeit with fewer analytical insights.

\section{Summary and Outlook}
\label{sec:conclusion}

In this paper, we have explored the application of mutual information (MI) as a tool for analyzing natural datasets, leveraging the strong interplay between machine learning and tensor network representations from quantum many-body theory. Our investigation uncovers several key insights into the structure of text and image data, with important implications for their efficient representation and processing.

For text data, we observed a power-law scaling of MI, suggesting that traditional one-dimensional tensor network approaches like matrix product states (MPS) and tree tensor networks (TTN) are not optimal for representing long texts. This contrasts with quantum systems, where power-law decaying correlations typically lead to logarithmic entanglement entropy scaling; hence MPS states provide efficient numerical descriptions. Our results indicate that classical text data exhibit a fundamentally different information structure, with power-law correlations coexisting with near-volume-law MI scaling. To better understand this phenomenon, we introduced a random pair model and an enhanced Markov generative model based on dependency parsing trees, which capture linguistic dependencies more accurately than the former. Both models successfully reproduce the observed mutual information scaling and the scaling of the correlation functions by carefully choosing the power-law distribution for the lengths of dependencies between words. This suggests that the hierarchical and statistical properties of natural language play a significant role in shaping its information structure.

For image data, our findings were more nuanced. For simpler datasets like MNIST~\cite{mnist}, we observed a clear area law scaling when the data were made translationally invariant. This result aligns well with previous successes in applying tensor network-based machine learning algorithms to MNIST classification tasks~\cite{Stoudenmire2016Supervised,Huang2017Provably,Glasser2019Expressive,Cheng2021Supervised}. However, for more complex image datasets such as Fashion-MNIST~\cite{Xiao2017Fashionmnist} and CIFAR-10~\cite{cifar}, the results were less definitive. While the MI for center-surrounding partitions adhered to an area law, the top-bottom partitions scaled more rapidly, indicating a deviation from area law scaling. This suggests that more sophisticated tensor network architectures~\cite{Glasser2018NeuralNetwork,Glasser2019Expressive} or hybrid models combining tensor networks with neural networks~\cite{Liu2023Tensor} might be necessary for effectively representing and processing these more complex image datasets. Further refinement and exploration of these hybrid models could pave the way for scalable tensor network applications in diverse machine learning tasks.

Our study underscores the potential of MI as a theoretical tool to guide the selection, improvement, and evaluation of machine learning models. Just as entanglement entropy and MI have played a crucial role in developing algorithms for quantum many-body physics, MI could serve a similar function in machine learning, aiding in capturing the necessary representation power to efficiently characterize complex datasets.

Several exciting avenues for future research emerge from our work. One promising direction is to characterize the MI scaling of distributions learned by state-of-the-art neural networks, such as gated recurrent units~\cite{Chung2014Empirical} and transformers~\cite{Vaswani2017Attention}. Such an analysis could provide insights into why contemporary language models, like BERT~\cite{Devlin2019BERT} and GPT~\cite{Brown2020Language}, outperform earlier models like recurrent neural networks~\cite{Sutskever2011Generating} and long short-term memory networks~\cite{LSTM} in generating coherent text over long sequences. Additionally, studying the dynamics of MI across the layers of deep learning networks, which has already offered new perspectives on their learning and information processing capabilities~\cite{Tishby2015Deep,Tishby2000Information,Goldfeld2018Estimating}, could be a fruitful area of future research. This could involve exploring how MI between data subregions evolves as neural networks process images or text sequentially. Another intriguing line of inquiry involves identifying other information-theoretic measures that could further our understanding of how datasets occupy only a small fraction of the entire parameter space and the traits that enable their efficient compression by tensor networks or analysis via neural networks. Advancements in these areas could enhance machine learning architectures and contribute to demystifying the inner workings of neural networks.

\begin{acknowledgments}
    We thank \'Alvaro M. Alhambra, Georgios Styliaris, and Samuel Scalet for fruitful discussions.
    This research is part of the Munich Quantum Valley, which is supported by the Bavarian state government with funds from the Hightech Agenda Bayern Plus.
    The work is partially supported by the Deutsche Forschungsgemeinschaft (DFG, German Research Foundation) under Germany's Excellence Strategy – EXC-2111 – 390814868.
    I.K.\ is supported by the Max-Planck-Harvard Research Center for Quantum Optics (MPHQ).
\end{acknowledgments}

%

\appendix
\renewcommand\thefigure{\thesection.\arabic{figure}}
\setcounter{figure}{0}

\section{More Details on the \texorpdfstring{$k$}{k}-Nearest Neighbor Estimator and Further Analysis}
\label{App:kNN}

Throughout this paper, we have utilized the well-known $k$-nearest neighbor (kNN) estimator~\cite{Kraskov2004Estimating} as a benchmark for other estimators. The kNN estimator is particularly effective for low-dimensional data. A key feature of this estimator is the additive constant in its mutual information (MI) estimates, dependent on the number of input data points, $n_\text{data}$, and the parameter $k$, representing the number of nearest neighbors used in the density estimation step (refer to Sec.~\ref{sec:MI_est} for more details).

\begin{figure*}
    \includegraphics[width=0.98\linewidth]{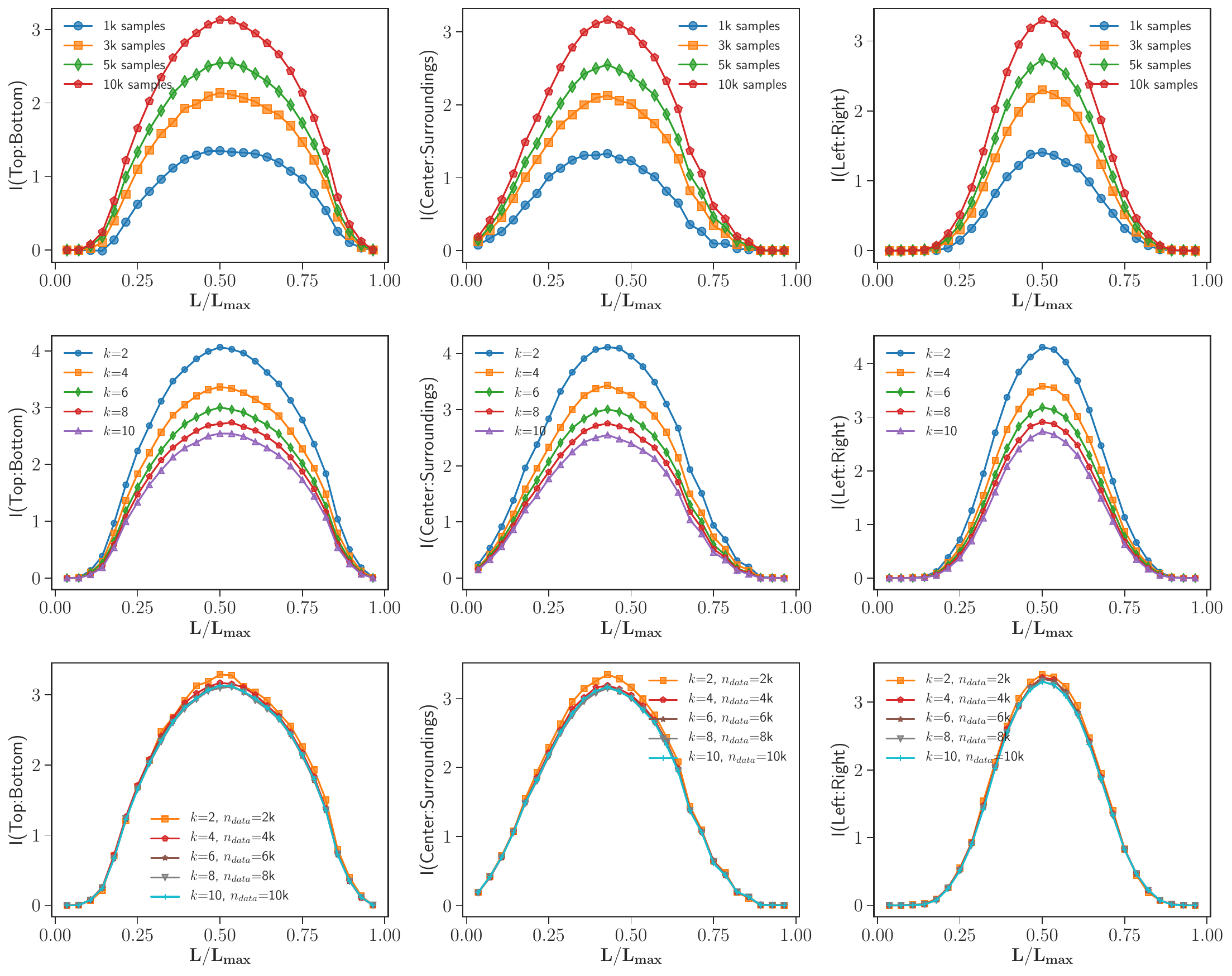}
    \caption{Detailed analysis of the kNN estimation of MI in the MNIST dataset.
    Top row, from Left to Right: Estimated $I(\text{L}:\text{R})$, $I(\text{C}:\text{S})$, $I(\text{T}:\text{B})$ with $k=10$ for 1k, 3k, 5k, 10k samples. Second row, from Left to Right: Estimated $I(\text{L}:\text{R})$, $I(\text{C}:\text{S})$, $I(\text{T}:\text{B})$ with 5k samples and $k=2,~4,~6,~8,~10$. The kNN estimation is influenced by the parameters $k$ and the number of samples used. Bottom row: Estimated $I(\text{L}:\text{R})$, $I(\text{C}:\text{S})$, $I(\text{T}:\text{B})$ with $k\propto{n_\text{data}}$. For $k\propto{n_\text{data}}$, all curves collapse, yielding a universal estimate. The $x$-axis represents the normalized length of the left, central, or top region, $L/L_{\max}$, and the $y$-axis shows the mutual information $I(A:B)$. These results demonstrate the consistency and scalability of the kNN estimator, as well as its sensitivity to dataset characteristics like edge effects.}
    \label{fig:kNN:test}
\end{figure*}

In Fig.~\ref{fig:kNN:test}, we present the MI estimates derived from the kNN method for the MNIST dataset without applying translational invariance to the images. Our empirical findings indicate that the scaling depends solely on the ratio $k/n_\text{data}$. Due to this dependency, we adjust the kNN results by a global additive constant when comparing with other methods unaffected by this issue.

Figure~\ref{fig:kNN:test} also illustrates the impact of the blank areas located at the edges of MNIST images, as the MI diminishes towards these edges. Moreover, the curves for the top:bottom partition do not exhibit the flat plateau observed in Fig.~\ref{fig:MI:images}(a), indicative of area law behavior in translationally invariant MNIST images. This deviation arises from the edges suppressing MI values, even near the center of the images.

\section{More Details on the Mutual Information Neural Estimator and Further Analysis}
\label{App:MINE}

In this appendix, we provide additional technical details on our implementation of the mutual information neural estimator (MINE)~\cite{Belghazi2018MINE}. This approach allows for the flexible use of various neural networks as score functions. As a lower-bound estimator, the accuracy of the MI estimation can be enhanced by achieving a higher MI estimate using a more expressive score function.

Initially, as per~\citet{Belghazi2018MINE}, we used a fully connected feedforward neural network (FC-FFNN) as the score function $T_{\theta}$. To leverage a network with greater expressive power and suitability for images and text, we switched to using convolutional neural networks (CNNs) as $T_{\theta}$. The CNN architecture consists of multiple layers: a 3D convolution layer, a 2D max-pooling layer, and a dropout layer~\cite{Srivastava2014Dropout} with a rate of $0.15$. This is followed by a fully connected flattening layer with $0.55$ dropout regularization. The last layer is then connected to the final fully connected layer with a single output. This output serves as the score function $T_{\theta}$ in Eq.~\eqref{eq:defMINE}. We employed the Adam optimizer~\cite{Kingma2015Adam,Reddi2018Convergence} with a batch size of 128 and a learning rate of $10^{-4}$ during training. The activation function used was ReLU~\cite{Nair2010Rectified}.

We compared the above-described CNN with an FC-FFNN, composed of an input layer matching the data dimensionality (e.g., $28\times28$ for MNIST), a hidden layer with $500$ neurons using ReLU activation, and a subsequent output layer with a single linear neuron. Both network types were trained under identical conditions regarding optimizer settings~(learning rate, batch size) and training epochs. The mutual information neural estimation procedure was implemented using PyTorch~\cite{Paszke2019PyTorch}.

\begin{figure*}
    \includegraphics[width=0.98\linewidth]{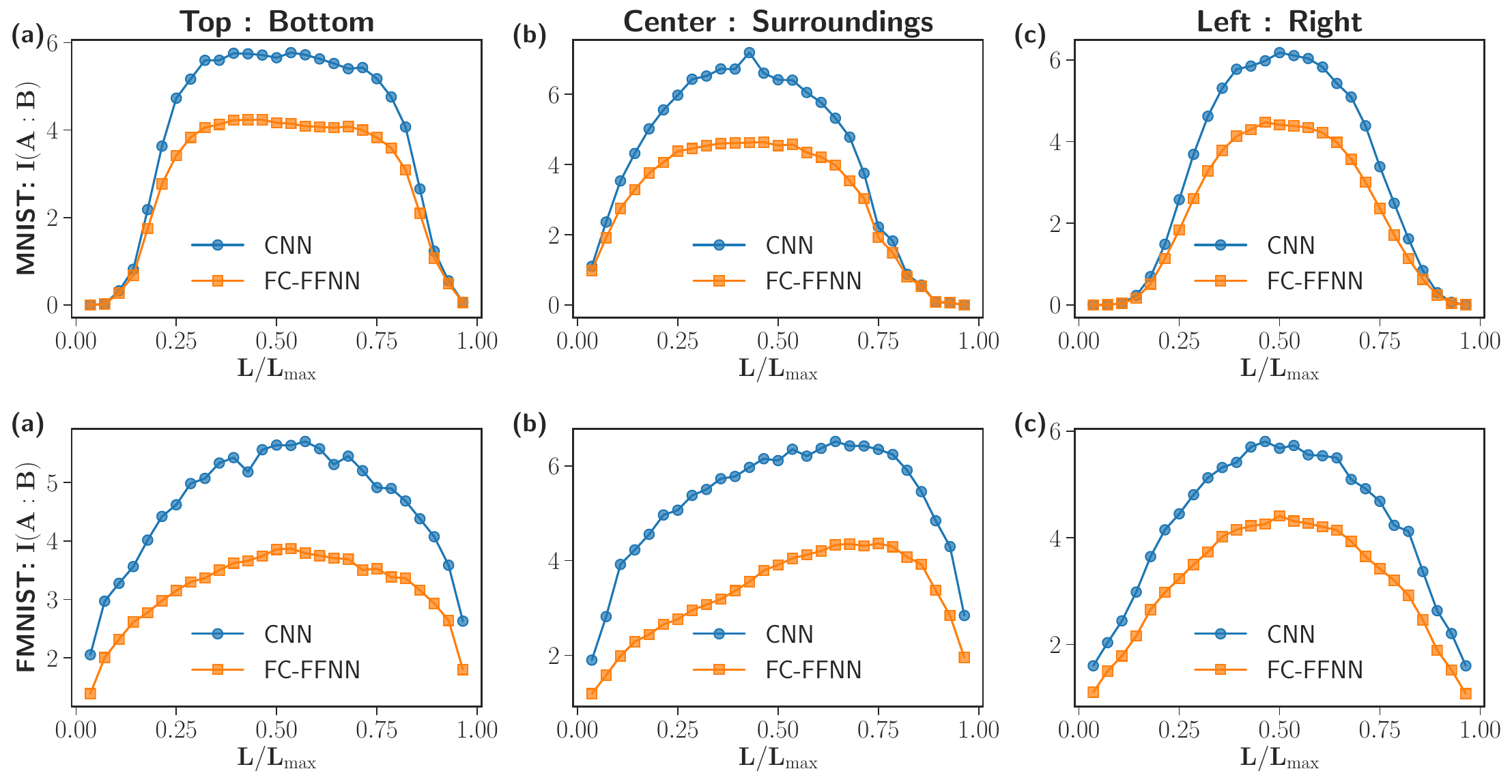}
    \caption{Detailed comparison of MINE estimation of MI in MNIST and Fashion-MNIST datasets using different score functions. Top row: $I(\text{L}:\text{R})$, $I(\text{C}:\text{S})$, $I(\text{T}:\text{B})$ estimates when $T_{\theta}$ is a convolutional neural network (CNN). Bottom row: $I(\text{L}:\text{R})$, $I(\text{C}:\text{S})$, $I(\text{T}:\text{B})$ estimates when $T_{\theta}$ is a fully connected feedforward neural network (FC-FFNN). The CNN consistently outperforms the FC-FFNN by providing higher MI estimates, indicating greater accuracy and stability. The $x$-axis represents the normalized length of the left, central, or top region, $L/L_{\max}$, and the $y$-axis shows the mutual information $I(A:B)$ in bits. These comparisons demonstrate the importance of choosing appropriate neural network architectures for accurate MI estimation in complex datasets.}
    \label{fig:MINE:FFNNvsCNN:MNIST}
\end{figure*}

In Fig.~\ref{fig:MINE:FFNNvsCNN:MNIST}, we compare the performance of MINE using FC-FFNN and CNN score functions on MNIST and Fashion-MNIST data. We found that CNNs consistently outperform FC-FFNNs by providing higher MI values. Consequently, we utilized CNN-based MINE results throughout the main text.

\clearpage

\section{Dependency Tree Model}
\label{App:DepTreeModel}

\begin{figure}[ht]
    \includegraphics[width=0.8\linewidth]{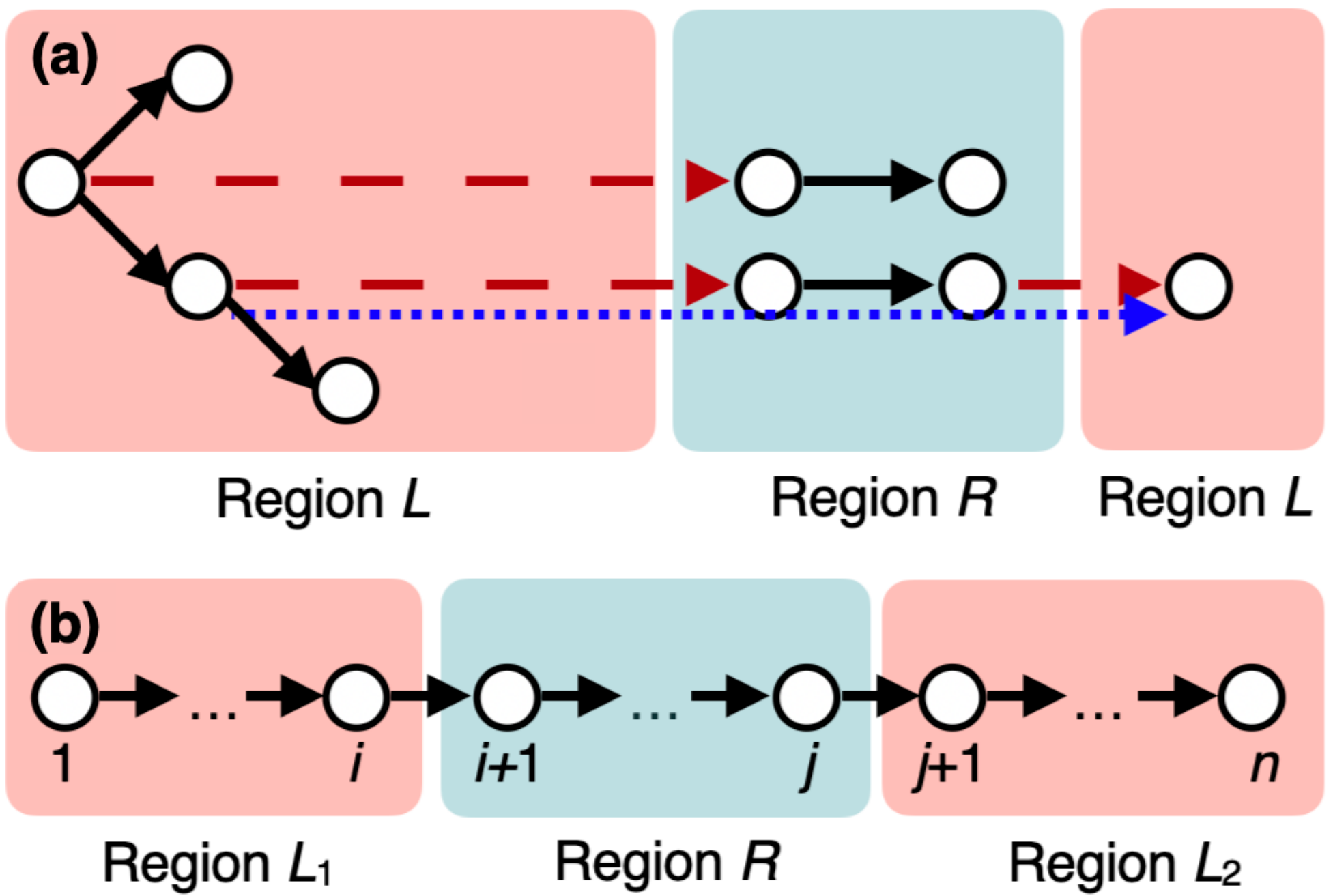}
    \caption{Markov chain model of text. (a)~The tree in Fig.~\ref{fig:Dependency_parsing_sec} expanded horizontally such that boundary crossings only occur from left to right. (b)~A simplified case of (a) when the tree consists of a single path.}
    \label{fig:Dependency_parsing_app}
\end{figure}

In this appendix, we provide detailed calculations for the dependency tree model introduced in Sec.~\ref{Sec:DepTreeModel}. First, we derive Eq.~\eqref{eq:MI_crossings_returns} in the simplified case of a linear dependency tree, as depicted in Fig.~\ref{fig:Dependency_parsing_app}~(a). This simplification is for notational convenience, and the derivations are similar in the general case.

In this linear tree of length $n$, the sequence starts with words in the left region, $L_1 = (W_1, W_2, \dots, W_i)$, crosses into the right region, $R = (W_{i+1}, W_{i+2}, \dots, W_j)$, and then returns to the left region with words $L_2 = (W_{j+1}, W_{j+2}, \dots, W_n)$. The mutual information between the left $L = L_1 \cup L_2$ and right $R$ regions is given by
\begin{align}
    I(L:R) &= S(L) + S(R) - S(L, R) \nonumber\\
           &= S(L_1, L_2) + S(R) - S(L_1, L_2, R) \nonumber\\
           &= \bigl[ S(L_1) + S(L_2) - I(L_1 : L_2) \bigr] \nonumber\\
           &\quad + S(R) - S(L_1, L_2, R), \label{eq:MI_Markov_chain}
\end{align}
where $I(L_1 : L_2)$ is the mutual information between $L_1$ and $L_2$.

Assuming the Markov property--that each word depends only on its parent--we can express the entropy of a sequence of words from $W_l$ to $W_m$ as
\begin{equation}
    S(W_l, \dots, W_m) = S(W_l) + \sum_{k=l}^{m-1} S\left(W_{k+1} \mid W_k\right).
\end{equation}
Applying this to the terms in Eq.~\eqref{eq:MI_Markov_chain} and simplifying, many terms cancel telescopically, resulting in:
\begin{equation}
    I(L:R) = I(W_{i+1} : W_i) + I(W_{j+1} : W_j) - I(L_1 : L_2).
\end{equation}
Here, $I(W_{i+1} : W_i)$ and $I(W_{j+1} : W_j)$ represent the MI contributions from the \emph{crossings} at the boundaries between $L$ and $R$ [sites $(i, i+1)$ and $(j, j+1)$]. The term $I(L_1 : L_2)$ accounts for the MI within the left region due to the \emph{returns}--paths that start and end in $L$ after passing through $R$, as we show below. The joint probability distribution of the words in $L_1$ and $L_2$ can be written as
\begin{equation}
    \mathbb{P}(L_1, L_2) = \mathbb{P}(L_1) \, \mathbb{P}(L_2) \, \frac{\mathbb{P}(W_{j+1}| W_i)}{\mathbb{P}(W_{j+1})},
\end{equation}
due to the Markov property. Consequently, the entropy of the two regions is
\begin{equation}
    S(L_1, L_2) = S(L_1) + S(L_2) - I(W_{j+1} : W_i).
\end{equation}
Hence, the mutual information between the left and right regions is given by the mutual information between the words at the boundary,
\begin{align}
    I(L_1 : L_2) &= S(L_1) + S(L_2) - S(L_1, L_2)\\
                 &= I(W_{j+1} : W_i),
\end{align}
which is a \emph{return} term.

\subsection{General Case: Branching Dependency Tree}

In the general case, where the dependency tree has multiple branches, as depicted in Fig.~\ref{fig:Dependency_parsing_app}~(b), the same reasoning applies independently to each branch connecting a parent word to one of its descendants. An analogous derivation leads to Eq.~\eqref{eq:MI_crossings_returns}. Assuming Markovian language generation, this formula is exact. This provides an opportunity to numerically benchmark various language models against empirical observations on real text, as in Sec.~\ref{sec:MI_classical_text}. This will be addressed in future work. In the remainder of this section, we aim to gain analytical understanding of the MI scaling based on the simplifying assumptions mentioned in Sec.~\ref{Sec:DepTreeModel}.

\subsection{MI Estimate Between Individual Words}

Let us denote the Markov matrix for word generation $\mathbf{M}$. We will make the standard assumption that $\mathbf{M}$ is both irreducible and aperiodic, thus it has a unique stationary distribution $\boldsymbol{\pi} = \mathbf{M} \, \boldsymbol{\pi}$. According to the Perron--Frobenius theorem, all other eigenvalues $\lambda_i$ satisfy $1 > |\lambda_2| \geq |\lambda_3| \geq \dotsb \geq 0$. With the left and right eigenvectors $\mathbf{l}_i$ and $\mathbf{r}_i$, $\mathbf{M}$ has the eigendecomposition:
\begin{equation}
    \mathbf{M} = \boldsymbol{\pi} \mathbf{1}^\mathrm{T} + \sum_{i=2}^m \lambda_i \mathbf{r}_i \mathbf{l}_i^\mathrm{T},
    \label{eq:MarkovEig}
\end{equation}
where $m$ is the dimensionality of the word space and $\mathbf{1} = (1, 1, \dots, 1)^\mathrm{T}$ is a constant column vector. Due to the biorthogonality between left and right eigenvectors $\mathbf{l}_i^\mathrm{T} \mathbf{r}_j = \delta_{ij}$, the $n$th power of the Markov matrix is
\begin{equation}
    \mathbf{M}^n = \boldsymbol{\pi} \mathbf{1}^\mathrm{T} + \sum_{i=2}^m \lambda_i^n \mathbf{r}_i \mathbf{l}_i^\mathrm{T},
\end{equation}
which decays exponentially with $n$, with the decay rate dominated by the second largest eigenvalue $\lambda_2$.

Next, let us estimate the mutual information between two words within the graph. Using Eq.~\eqref{eq:MI_crossings_returns}, we relied on this result to estimate MI scaling between text regions in Sec.~\ref{Sec:DepTreeModel}.

The crossing terms in Eq.~\eqref{eq:MI_crossings_returns} are between a parent word $W_p$ and its child word $W_c$ in the tree structure where the distribution of $W_p$ is given by $\boldsymbol{\mu}$. The joint distribution of $W_p$ and $W_c$ is given by $\mathbb{P}(W_p = b, W_c = a) = \mu_b M_{ba}$, where $M_{ba} = \mathbb{P}(W_c = a \,|\, W_p = b)$. The marginal distribution of $W_c$ is $\mathbb{P}(W_c = a) = (\mathbf{M}^\mathrm{T} \boldsymbol{\mu})_a$. Thus, the mutual information is
\begin{equation}
    I(W_p : W_c) = \sum_{a,b} \mu_b M_{ba} \log\left( \frac{M_{ba}}{(\mathbf{M}^\mathrm{T} \boldsymbol{\mu})_a} \right).
\end{equation}
In our approximations, we will assume that $\boldsymbol{\mu}$ is close to the stationary distribution $\boldsymbol{\mu} \approx \boldsymbol{\pi}$. Thus, the mutual information is approximately constant for \emph{crossings} in the tree graph:
\begin{equation}
    I(\text{crossing}) \approx \mathcal{C}_c.
    \label{eq:crossing_constant_MI}
\end{equation}
Assume that words $U$ and $V$, with word distribution vectors $\boldsymbol{\mu}_U$ and $\boldsymbol{\mu}_V$, share an earliest common ancestor $X$ in the tree, with distances $\Delta_U$ and $\Delta_V$ to $U$ and $V$, respectively. Similarly to the previous linear case, we assume that the distribution of $X$ is close to the stationary one $\boldsymbol{\mu}_X \approx \boldsymbol{\pi}$.
To derive a similar formula for \emph{returns}, we use much of the notation from Ref.~\onlinecite{Lin2017Critical} and their assumption that the distance between words is large enough so that the words are approximately independent from each other,
\begin{equation}
    \mathbb{P}(U = u, V = v) \approx \mathbb{P}(U = u) \, \mathbb{P}(V = v). \label{eq:approx_independence}
\end{equation}
The mutual information can thus be approximated as~\cite{Lin2017Critical}
\begin{align}
    I(U:V) &= \sum_{u,v} \mathbb{P}(u, v) \log \left( \frac{\mathbb{P}(u, v)}{\mathbb{P}(u) \mathbb{P}(v)} \right) \nonumber\\
           &= \sum_{u,v} \mathbb{P}(u, v) \log \left( 1 + \frac{\mathbb{P}(u, v)}{\mathbb{P}(u) \mathbb{P}(v)} - 1 \right) \nonumber\\
           &\lesssim \sum_{u,v} \mathbb{P}(u, v) \left( \frac{\mathbb{P}(u, v)}{\mathbb{P}(u) \mathbb{P}(v)} - 1 \right) \nonumber\\
           &= \sum_{u,v} \frac{\mathbb{P}(u, v)^2}{\mathbb{P}(u) \mathbb{P}(v)} - 1.
    \label{eq:MI_large_distance}
\end{align}
Note that this is a strict upper bound due to Jensen's inequality ($\log(1 + x) \leq x$ for $x \geq 0$), which is a good approximation if Eq.~\eqref{eq:approx_independence} holds.
The joint probability distribution can be further approximated up to leading order in the eigenvalues of $\mathbf{M}$ as
\begin{align*}
    \mathbb{P}(u, v) & \approx \sum_{x} \pi_{x} \left( \pi_{u} + \lambda_2^{\Delta_U} A_{x u} \right) \left( \pi_{v} + \lambda_2^{\Delta_V} A_{x v} \right) \\
                     & = \pi_{u} \pi_{v} + \lambda_2^{\Delta_U + \Delta_V} \sum_{x} \pi_x A_{x u} A_{x v},
\end{align*}
where the matrix $\mathbf{A} = \mathbf{r}_2 \, \mathbf{l}_2^{T}$ is the projector to the subspace with the second largest eigenvalue. $A_{x u}$ captures the influence of ancestor $X=x$ on word $U=u$. In the formula above, the cross terms vanish for this reason, since $\pi_x$ belongs to the subspace of $\lambda_1 = 1$. Similarly, in Eq.~\eqref{eq:MI_large_distance}, the cross terms in $\mathbb{P}^2(u,v)$ vanish since $\sum_u A_{xu}\pi_u = 0$.
Hence, up to the same order, the mutual information becomes
\begin{equation}
    I(U:V) \approx \left( \sum_{u,v} \frac{\left( \sum_{x} \pi_x A_{x u} A_{x v} \right)^2}{\pi_u \pi_v} \right) \lambda_2^{2 (\Delta_U + \Delta_V)}.
\end{equation}
This expression indicates that the MI between two words $U$ and $V$ decays exponentially with their combined distances from their earliest common ancestor in the dependency tree. Therefore, the \emph{return} contributions in Eq.~\eqref{eq:MI_crossings_returns} scale with $\text{dist}$, the number of edges between them in the graph, as
\begin{equation}
    I(\text{return}) \approx \mathcal{C}_r \cdot \lambda_2^{2\, \text{dist}} + \mathcal{O}\left( \left| \frac{\lambda_3}{\lambda_2} \right|^{2\, \text{dist}} \right).
    \label{eq:return_MI_scaling}
\end{equation}

\end{document}